\documentclass[aps,pra,twocolumn,groupedaddress,nofootinbib,showpacs]{revtex4-1}
\usepackage{amsmath}
\usepackage{mathrsfs}
\usepackage[pdftex,colorlinks=true,urlcolor=blue,linkcolor=blue,citecolor=blue]{hyperref}
\usepackage{graphicx}
\usepackage{float}
\usepackage[abs]{overpic}
\usepackage{color}
\usepackage{subfigure}
\begin{document}

\title{Kinetics of Bose-Einstein condensation in a dimple potential}

\author{Shovan Dutta}
\email{sd632@cornell.edu}
\author{Erich J. Mueller}
\email{em256@cornell.edu}
\affiliation{Laboratory of Atomic and Solid State Physics, Cornell University, Ithaca, New York 14850, USA}

\date{\today}

\begin{abstract}
We model the dynamics of condensation in a bimodal trap, consisting of a large reservoir region, and a tight ``dimple" whose depth can be controlled. Experimental investigations have found that such dimple traps provide an efficient means of achieving condensation. In our kinetic equations, we include two- and three-body processes. The two-body processes populate the dimple, and lead to loss when one of the colliding atoms is ejected from the trap. The three-body processes produce heating and loss. We explain the principal trends, give a detailed description of the dynamics, and provide quantitative predictions for timescales and condensate yields. From these simulations, we extract optimal parameters for future experiments.
\end{abstract}


\maketitle

\section{Introduction}\label{intro}

Cold atom physics faces a crisis. Experimentalists are trying to produce ever-more complicated states of matter which are intrinsically difficult to cool \cite{demarco11, bloch08, mueller08, ho09, ho07}. To further complicate matters, the techniques used to produce these states add mechanisms for heating. One paradigm for addressing this crisis is to divide the system into two parts: a ``reservoir" which can readily be cooled, coupled to a smaller subsystem which has interesting properties \cite{ketterle98dimple, pinkse97, grimm03, dalibard11, davis11, grimm13, bernier09, hoarxiv, dagotto09, kollathrecent}. The prototypical example of such a separation are the ``dimple traps" pioneered by Pinkse {\it et al.} \cite{pinkse97}, and more recently explored by several other groups \cite{ketterle98dimple, grimm03, dalibard11, davis11, grimm13}. The key to such programs is an understanding of the kinetic processes through which energy and particles move between the two subsystems. Here we model the loading and equilibration of a dimple trap in a gas of weakly interacting Bosons.

The initial dimple experiments were motivated by a desire to reduce the complexity of cooling atomic gases, and explore fundamental questions of condensate growth \cite{pillet06, cornell95, hulet95, ketterle95, ketterle98evaporative, cornell01, esslinger02, sengstock04, esslinger07, walraven96, ketterle95theory, gardiner97trap, gardiner98, gardiner00constantbath, gardiner00changingbath, stoof97, holland97, gardiner97notrap, holland99, stoof00}. Dimples have been key to proposals to study atom lasers \cite{ketterle97, bloch99, grimm13}. They also have promising applications in atom interferometry \cite{pritchard09}, quantum tweezers for atoms \cite{niu02}, controlling soliton-sound interaction \cite{adams03}, ultraslow light propagation \cite{uncu12}, and studying analogs of cosmological physics \cite{kz}.

Stellmer {\it et al.} describes a typical dimple experiment in Ref. \cite{expdetails}. They precool a cloud of Bosonic atoms to hundreds of nanokelvin, and trap them in a large but shallow optical trap. At this point, the phase space density is well below the threshold for condensation. Next a laser beam focused in a small region near the trap center creates a strongly attractive dimple potential, causing a great increase in the local atom density without much change in temperature. As the density in the dimple grows, they see the development of a condensate. Theoretical steps have been taken to understand the kinetics of Bose condensation by this method \cite{davis11, pillet06, stoof01, ma04, stoof06, uncu0708, uncu13linear, uncu13parabolic}. However, a detailed quantitative study of how the condensate fraction, the temperature, and the characteristic time scales depend on the trap parameters and the initial conditions is, to our knowledge, still lacking. Here we take a simple quantum kinetic approach towards achieving this goal.

Since in experiments like Stellmer {\it et al.}'s, the phase space density in the reservoir stays small, we model it as a classical Boltzmann gas in a three-dimensional (3D) harmonic well. We assume that the collision rate in the reservoir is fast compared to the condensation dynamics so that the reservoir is always in quasi-thermal equilibrium: the occupation of a mode of energy $\varepsilon$ is $n_{\varepsilon}^{r} (t) = e^{-(\varepsilon - \mu_r (t))/k_B T_r (t)}$, where the temperature $T_r$ and the chemical potential $\mu_r$ depend on time. Many cold atom experiments are described by such a quasi-equilibrium \cite{ketterle98dimple, pinkse97, davis11}. For simplicity we assume that the harmonic well is isotropic. We model the dimple as a 3D square well, and consider the case where it is turned on suddenly at $t = 0$. This is a prototypical protocol for turning on the dimple \cite{davis11, pillet06, ketterle98evaporative, stoof06}. In Sec. \ref{parameters}, we outline the physical parameters relevant to the dynamics. In Sec. \ref{kinetics model}, we analyze the two-body scattering processes responsible for the transfer of atoms from the reservoir to the dimple and their redistribution among the momentum states of the dimple. In particular, the dimple is populated via two-body collisions: one particle enters the dimple, transferring energy to the second. This permits us to write down rate equations for the populations of the dimple states. Due to the symmetry of kinetic processes, the population of a dimple state depends only on its energy. Since in most present-day experiments the dimple contains thousands of energy levels \cite{davis11, grimm13}, we describe their populations by a continuous distribution function $f(E,t)$, treating the ground state occupation ($N_0$) separately. As the collision processes also change the number of atoms ($N_r$) and the energy ($E_r$) in the reservoir, we arrive at coupled rate equations for $f(E,t)$, $N_0$, $N_r$, and $E_r$.

In Sec. \ref{thermalization section}, we discuss the short-time dynamics after turning on the dimple. We find that energy levels near half the dimple depth start filling up first. When the atom density in the dimple becomes comparable to that in the reservoir, particle scattering between energy levels initiates thermalization. Denoting the collisional mean free time in the reservoir by $\tau_{\mbox{\scriptsize{coll}}}$, we find that for $t \gtrsim 8 \tau_{\mbox{\scriptsize{coll}}}$, $f(E,t)$ is well approximated by a thermal distribution. The states in the high-energy tail take longer to thermalize than those near the bottom of the dimple. Similar thermalization time scales were reported in previous studies in a wide range of geometries \cite{ketterle98evaporative, esslinger07, walraven96, gardiner97trap, expdetails, wolfe89, kagan92}. During thermalization, for large enough dimple depths, we notice that $f(E,t)$ passes through a bimodal shape, which should show up in time-of-flight images in experiments. We find that $N_0 (t)$ grows slowly at first until it becomes sufficiently large that Bose stimulation takes over. This gives rise to an onset time $\tau_{\mbox{\scriptsize{on}}}$ after which the condensate grows rapidly. This effect was studied for condensation by evaporative cooling in harmonic traps \cite{ketterle98evaporative, gardiner97trap, gardiner98, gardiner00constantbath}, and has been observed in recent experiments on dimples \cite{davis11, grimm13}.

We consider both the cases of an infinite trap depth and a finite trap depth. In the former situation we allow particles in the reservoir to have arbitrarily high energies, whereas in the latter, we eject particles which recoil from a collision with an energy greater than the trap depth. In Sec. \ref{infinite trap depth}, we discuss the infinite trap depth case. In practice, this is equivalent to a trap whose depth is large compared to the dimple depth and the initial thermal energy. The reservoir temperature rises as particles are scattered into the dimple. In the absence of inelastic losses, the dimple population grows monotonically, saturating after a time $\tau_{\mbox{\scriptsize{sat}}}$ at a value limited by the amount of heating. Guided by the result that the dimple thermalizes fast compared to the population growth rates, we introduce a simplified model where we assume that $f(E,t)$ is given by a thermal Bose-Einstein distribution. This method reproduces all features of the full model for $t \gtrsim 10 \tau_{\mbox{\scriptsize{coll}}}$, and requires fewer computational resources to simulate. We only use this approximation in Sec. \ref{infinite trap depth}, returning to the full kinetic equations in later sections. We provide detailed results of how the final populations, the final temperature, the entropy gain, $\tau_{\mbox{\scriptsize{on}}}$, and $\tau_{\mbox{\scriptsize{sat}}}$ vary with the dimple depth $\varepsilon_d$, the initial phase space density $\rho_i$, and the ratio of the reservoir volume to the dimple volume $\Omega$, and the initial temperature $T_{r0}$. In particular, the atoms do not condense if $\varepsilon_d$ is smaller than $|\mu_r (t=0)|$. As $\varepsilon_d$ is increased, the final condensate fraction $F_0$ grows and attains a maximum for an optimal depth $\varepsilon_d^*$ which is set by $\rho_i$, $\Omega$, and $T_{r0}$. With further increase in $\varepsilon_d$, $F_0$ falls off due to increased heating. Such a non-monotonic variation was observed in a recent experiment \cite{davis11}. In addition to maximizing $F_0$, $\varepsilon_d = \varepsilon_d^*$ also minimizes $\tau_{\mbox{\scriptsize{on}}}$. Both $F_0$ and $\varepsilon_d^*$ increase with $\rho_i$ and $\Omega$. We find that $\Omega \tau_{\mbox{\scriptsize{coll}}}$ sets the typical timescale for saturation. The dynamics become more non-adiabatic and takes longer to saturate at larger dimple depths.

We add inelastic losses to our model in Sec. \ref{threebody}. Here we consider the case for ${}^{87}$Rb where three-body recombination dominates the loss \cite{dalibard99}. In this process three atoms collide to produce a molecule in an excited state, thereby releasing  a large amount of energy which causes all three atoms to escape. As a result, the condensate fraction decays toward zero after reaching a peak value $F_0^{\mbox{\scriptsize{peak}}}$ at $t = \tau_{\mbox{\scriptsize{peak}}}$. Thus three-body loss gives a finite condensate lifetime $\Delta t_{\mbox{\scriptsize{lf}}}$ \cite{grimm13}. We find that $F_0^{\mbox{\scriptsize{peak}}}$ exhibits a non-monotonic dependence on $\varepsilon_d$ similar to $F_0$. However, the maximum condensate fraction is smaller by almost an order of magnitude due to the large three-body loss rate in the dimple where the density becomes large. We find that three-body loss also lowers the optimal dimple depth, in agreement with recent findings \cite{davis11}. Smaller dimples result in higher local densities, which increase the loss rate. We therefore find an optimal volume ratio $\Omega^*$. Similarly, there is an optimal initial phase space density $\rho_i^*$. The three-body rate grows faster with $T_{r0}$ than the two-body collision rate. Thus $F_0^{\mbox{\scriptsize{peak}}}$ falls off with $T_{r0}$. We find that $\Delta t_{\mbox{\scriptsize{lf}}}$ increases with $\Omega$, and decreases with $\varepsilon_d$, $\rho_i$, and $T_{r0}$. Since the three-body loss rate varies with the s-wave scattering length $a$ as $|a|^4$ \cite{esry99}, one can influence it by using a different species of atoms or exploiting Feshbach resonances \cite{grimm03}. However, such manipulations may introduce other inelastic channels, and one should be careful to take that into account.

In Sec. \ref{finite trap depth}, we discuss how our results change when the reservoir trap has a finite depth $\varepsilon_t$. Here we eject any atom which gains sufficient energy from a collision to have a total energy $\varepsilon > \varepsilon_t$. Such a model correctly describes a trapped gas in the Knudsen regime: the collisional mean free path is larger than the size of the reservoir, which is true in most experiments on trapped gases \cite{walraven96}. We assume that the atom energies in the reservoir follow a Boltzmann distribution truncated at $\varepsilon = \varepsilon_t$. Previous numerical studies have shown that this assumption accurately describes evaporative cooling \cite{walraven96, ketterle95theory,gardiner97trap}. The effect of finite trap depth on the dynamics becomes appreciable when $\varepsilon_t$ is no longer larger than $\varepsilon_d$ and $k_B T_{r0}$. We find that lower trap depths yield lower final temperatures, and increase the condensate growth rate. This leads to a higher condensate fraction $F_0^{\mbox{\scriptsize{peak}}}$ and a longer lifetime $\Delta t_{\mbox{\scriptsize{lf}}}$. However, when $\varepsilon_t$ becomes very small, the increased evaporation rate of the reservoir limits the rise of $\Delta t_{\mbox{\scriptsize{lf}}}$. We summarize our findings and suggest future work in Sec. \ref{summary}.

To keep the problem computationally tractable, we have made some simplifying approximations. First, we have not included the mean-field interactions between the condensate and the thermal cloud \cite{davis11, stoof01, stoof00, gardiner98, gardiner00constantbath, gardiner00changingbath, holland99, aspect04, cornell97, dalfovo99}. This mean-field changes the effective potential experienced by the atoms by an amount proportional to the condensate density, in effect changing the dimple depth. This causes a repulsion between the condensate and thermal cloud \cite{gardiner00constantbath, aspect04}, and also lowers the critical temperature \cite{cornell97}. It can be compensated by making the dimple parameters time-dependent, and is unimportant for the short-time dynamics. We do not model quantum fluctuations of the condensate \cite{cornell97, stoof01, holland99, gardiner97trap}, although our model does include thermal fluctuations. The modifications due to quantum fluctuations should be much smaller than that of the mean-field \cite{gardiner97trap, cornell97}. We use the infinite square well energy eigenstates for the dimple. The kinetics do not depend on the exact model of the dimple potential as long as it contains many energy levels, which is true for present-day experiments \cite{davis11, grimm13}. In modeling the condensation kinetics, we neglect two kinds of elastic collisions: First, we neglect collisions internal to the dimple where there is no exchange of atoms with the reservoir. These processes serve to equilibrate the dimple. Within our approximations, we find that the dimple thermalizes within $\tau_{\mbox{\scriptsize{th}}} \approx 8 \tau_{\mbox{\scriptsize{coll}}}$, and these processes can at most speed up thermalization. For $t > \tau_{\mbox{\scriptsize{th}}}$, these neglected collisions play no role. Second, we neglect collisions in which an atom from a low-energy state in the dimple and an atom from the reservoir collide, leaving two atoms in the dimple. Such collisions can become important for deep dimples when the condensate fraction becomes appreciable, and can subsequently enhance the dimple population rate. They can be included in a future refinement of our model, and might alter some quantitative details such as the time to reach saturation. However, we do not expect them to change any of the qualitative features our model captures \cite{gardiner98}.

\section{Formalism}\label{formalism}

\subsection{Physical parameters of the dimple potential}\label{parameters}

In this subsection we describe our model for the reservoir and the dimple, and develop some useful notation.

We model the reservoir as an isotropic harmonic well of frequency $\omega$, truncated at the trap depth $\varepsilon_t$, and assume a truncated Boltzmann distribution. We can relate the number of atoms $N_r$ and the energy $E_r$ in the reservoir to $\varepsilon_t$, the inverse temperature $\beta_r = 1 / k_B T_r$, and the fugacity $z_r = e^{\beta_r \mu_r}$ by integrating over phase space:
\begin{equation}
N_r = \int^{\prime} \frac{d^3 p d^3 r}{h^3} \;\exp\Big[-\beta_r \Big(\frac{p^2}{2 m} + \frac{1}{2} m \omega^2 r^2 - \mu_r\Big)\Big].
\label{number1}
\end{equation}
Here $m$ denotes the mass of an atom. The prime stands for the condition that any atom in the reservoir has a total energy less than $\varepsilon_t$, i.e., $\frac{p^2}{2 m} + \frac{1}{2} m \omega^2 r^2 < \varepsilon_t$. Eq. (\ref{number1}) can be simplified to obtain
\begin{equation}
N_r = \frac{z_r}{(\beta_r \hbar \omega)^3} \frac{1}{2} \gamma(3,\beta_r \varepsilon_t)\;,
\label{number2}
\end{equation}
where $\gamma$ denotes the lower incomplete gamma function. Similarly we find for the energy,
\begin{align}
\nonumber E_r &= \int^{\prime} \frac{d^3 p d^3 r}{h^3} \Big(\frac{p^2}{2 m} + \frac{1}{2} m \omega^2 r^2 \Big) \; e^{-\beta_r (\frac{p^2}{2 m} + \frac{1}{2} m \omega^2 r^2 - \mu_r)}\\
&= \frac{1}{\beta_r} \frac{z_r}{(\beta_r \hbar \omega)^3} \frac{1}{2} \gamma(4,\beta_r \varepsilon_t)\;. \label{energy}
\end{align}

After turning on the dimple, $N_r$, $E_r$, $\beta_r$, and $z_r$ change as functions of time. We define $f_r (t)$ and $e_r (t)$ as the ratio of $N_r (t)$ and $E_r (t)$ to their initial values, $\mathcal{N}$ and $\mathcal{E}$ respectively. Thus,
\begin{eqnarray}
f_r \equiv \frac{N_r}{\mathcal{N}} = \frac{\tilde{z}_r}{\tilde{\beta}_r^3} \frac{\gamma(3, \tilde{\beta}_r \tilde{\varepsilon}_t)}{\gamma(3, \tilde{\varepsilon}_t)}\;,\label{frdef}\\
e_r \equiv \frac{E_r}{\mathcal{E}} = \frac{\tilde{z}_r}{\tilde{\beta}_r^4} \frac{\gamma(4, \tilde{\beta}_r \tilde{\varepsilon}_t)}{\gamma(4, \tilde{\varepsilon}_t)}\;\label{er},
\end{eqnarray}
where $\tilde{z}_r \equiv z_r / z_{r0}$, $\tilde{\beta}_r \equiv \beta_r / \beta_{r0}$, and $\tilde{\varepsilon}_t \equiv \beta_{r0} \varepsilon_t$. The zeros in the subscripts refer to the respective values at $t=0$. In our simulation of the kinetics, we use Eqs. (\ref{frdef}) and (\ref{er}) to extract the instantaneous values of $\tilde{z}_r$ and $\tilde{\beta}_r$ from a knowledge of $f_r$ and $e_r$.

The spatial density of atoms in the reservoir can be found as
\begin{align}
\nonumber n_r (\vec{r}) &= \int^{\prime} \frac{d^3 p}{h^3} \;\exp\Big[-\beta_r \Big(\frac{p^2}{2 m} + \frac{1}{2} m \omega^2 r^2 - \mu_r\Big)\Big]\\
&= \frac{z_r}{\lambda_r^3} \frac{2}{\sqrt{\pi}} \gamma\Big(\frac{3}{2}, \beta_r \varepsilon_t - \frac{1}{2} \beta_r m \omega^2 r^2\Big) \; e^{-\frac{1}{2} \beta_r m \omega^2 r^2}.
\label{nr}
\end{align}
Here $\lambda_r = (2 \pi \hbar^2 \beta_r / m)^{1/2}$ denotes the thermal de Broglie wavelength. We see that $n_r$ and hence the phase space density falls off with distance from the center of the well. We define the ``initial phase space density" $\rho_i$ to be
\begin{equation}
\rho_i \equiv n_{r0} (\vec{0}) \lambda_{r0}^3 = z_{r0}\; \frac{2}{\sqrt{\pi}} \gamma(3/2, \tilde{\varepsilon}_t)\;.
\label{rhoi}
\end{equation}
This corresponds to the phase space density near the dimple at $r=0$.

The ``collisional mean free time" $\tau_{\mbox{\scriptsize{coll}}}$ is the average time between successive collisions among the atoms in the reservoir near $r=0$. We can estimate $\tau_{\mbox{\scriptsize{coll}}}$ at $t=0$ as $\tau_{\mbox{\scriptsize{coll}}} = (n \sigma v)^{-1}$, where $n = n_{r0} (\vec{0})$, $\sigma = 8 \pi a^2$ is the scattering cross section for weakly interacting Bosons \cite{pethick}, and $v$ denotes the average initial speed of the reservoir atoms near $r=0$. One can find $v$ as
\begin{align}
\nonumber v &= \frac{1}{n_{r0}(\vec{0})} \int^{\prime} \frac{d^3 p}{h^3} \frac{p}{m}\exp\Big[-\beta_{r0} \Big(\frac{p^2}{2 m} - \mu_{r0}\Big)\Big]\\
&= \Big(\frac{8}{\pi m \beta_{r0}}\Big)^{1/2} \frac{\sqrt{\pi}/2}{\gamma(3/2,\tilde{\varepsilon}_t)} \gamma(2,\tilde{\varepsilon}_t)\;.
\label{v}
\end{align}
Therefore
\begin{equation}
\tau_{\mbox{\scriptsize{coll}}} = \frac{(m \beta_{r0})^{1/2}}{16 \sqrt{2 \pi} a^2 n_{r0}(\vec{0})} \frac{2}{\sqrt{\pi}} \frac{\gamma(3/2,\tilde{\varepsilon}_t)}{\gamma(2,\tilde{\varepsilon}_t)}\;.
\label{taucoll}
\end{equation}

We define an effective volume $V_r$ of the reservoir as $V_r \equiv N_r / n_r (\vec{0})$. At $t=0$, this has the value
\begin{equation}
V_{r0} = \mathcal{N} / n_{r0} (\vec{0}) = \Big(\frac{2 \pi}{\beta_0 m \omega^2}\Big)^{3/2} \frac{\sqrt{\pi}}{4} \frac{\gamma(3, \tilde{\varepsilon}_t)}{\gamma(3/2, \tilde{\varepsilon}_t)}\;,
\label{Vr0}
\end{equation}
where we have substituted from Eqs. (\ref{number2}) and (\ref{nr}). The incomplete gamma functions become quite insensitive to $\tilde{\varepsilon}_t$ for $\tilde{\varepsilon}_t \gtrsim 5$ where $\gamma(\nu,\tilde{\varepsilon}_t) \approx \Gamma(\nu)$. Then $\rho_i$ and $V_{r0}$ are just functions of the trap frequency, the initial temperature, and the total number of trapped atoms. When $\tilde{\varepsilon}_t \lesssim 1$, the truncated Boltzmann distribution may no longer be a good model for the distribution.

We model the dimple as a square well of depth $\varepsilon_d$ and length $l_d$. We find that the condensation dynamics depends on the ratio of $V_{r0}$ to the dimple volume $l_d^3$. Thus we define the ``volume ratio" $\Omega \equiv V_{r0} / l_d^3$. Using Eqs. (\ref{rhoi}) and (\ref{Vr0}), one can write
\begin{equation}
\Omega \equiv \frac{V_{r0}}{l_d^3} = \frac{\mathcal{N} \lambda_{r0}^3}{l_d^3 \; \rho_i} = \frac{\mathcal{N}}{\tilde{l}_d^3} \frac{1}{z_{r0}} \frac{\sqrt{\pi}/2}{\gamma(3/2,\tilde{\varepsilon}_t)}\;,
\label{omega}
\end{equation}
where $\tilde{l}_d \equiv l_d / \lambda_{r0}$.

We assume that the eigenstates of the dimple coincide with those for the ``particle in a box" model, i.e., they are plane wave states of definite momenta:
\begin{equation}
\psi_{\vec{n}} (\vec{r}) = l_d^{-3/2}\hspace{0.03cm} e^{i \frac{2 \pi}{l_d} \vec{n}.\vec{r}}\;,
\label{psin}
\end{equation}
where $\vec{n}$ is a triplet of integers. Such a state has energy
\begin{equation}
\varepsilon_{n} \equiv - \varepsilon_d + E_{n} = - \varepsilon_d + \frac{2 \pi^2 \hbar^2}{m l_d^2} n^2\;,
\label{energyn}
\end{equation}
with $n = (n_1^2 + n_2^2 + n_3^2)^{1/2}$. We can estimate the total number of such states, $M$, by applying the condition that $\varepsilon_{n}$ must be negative. This gives $M \approx (1/6 \pi^2) (2 m l_d^2 \varepsilon_d / \hbar^2)^{3/2}$. For typical magnitudes of $\varepsilon$ and $l_d$ in present-day experiments, $M$ is very large \cite{davis11, expdetails}. Multiplying Eq. (\ref{energyn}) by $\beta_{r0}$, we can express it in the tilde notation as
\begin{equation}
\tilde{\varepsilon}_{n} \equiv - \tilde{\varepsilon}_d + \tilde{E}_{n} = - \tilde{\varepsilon}_d + (\pi/\tilde{l}_d^2) n^2\;.
\label{energyntilde}
\end{equation}
Other models for the dimple (such as a harmonic oscillator) yield similar results for the dynamics.

\subsection{Kinetic model for condensation in the dimple}\label{kinetics model}

\begin{figure}[t]
\includegraphics[width=\columnwidth]{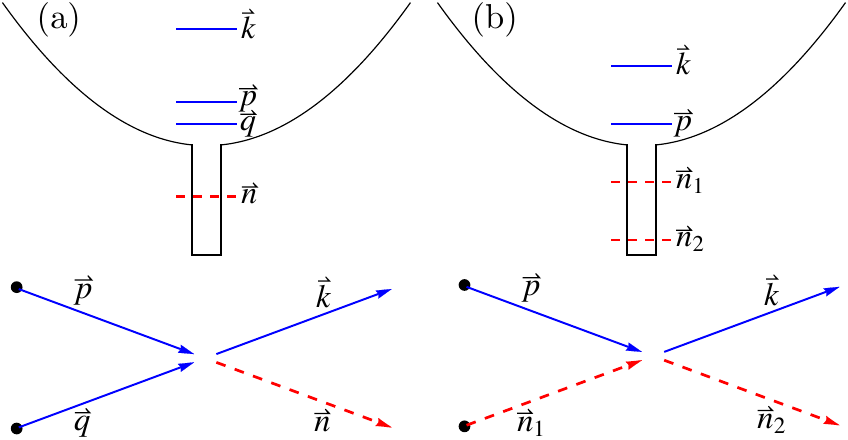}
\caption{(Color online) Two body collisions responsible for the growth and redistribution of the dimple population: (a) Two atoms from the reservoir collide and one of them enters a dimple state, transferring energy to the other. (b) A reservoir atom exchanges energy with an atom in the dimple, transferring it to a different energy state.}
\label{processes}
\vspace{-0.2cm}
\end{figure}

To model the non-equilibrium dynamics after the dimple is turned on, we consider the two different kinds of two-body elastic collisions which dominate the energy and particle transport between the reservoir and the dimple. These are illustrated in Fig. \ref{processes}. In the first kind, a collision between two atoms in the reservoir transfers one of the atoms to the dimple, while the other atom gains energy. The second atom can leave the reservoir if its total energy exceeds the trap depth $\varepsilon_t$. However, when $\varepsilon_t$ is large compared to $\varepsilon_d$, it is more likely that the second atom will stay in the reservoir and cause heating. Collisions of this kind lead to the growth of the dimple population, and increase the local phase space density.
In the second kind of collision, an atom in the reservoir collides with an atom in the dimple, and transfers it to another energy state in the dimple. The first atom can then either remain in the reservoir or leave, depending on the amount of energy it gains or loses in the process. Collisions of this second kind serve to thermalize the dimple by redistributing its atom population among the various energy levels. In the following, we analyze these two kinds of collisions (and their reverse processes) in detail to derive the equations of motion for the dynamics.
\vspace{-0.3cm}
\subsubsection{Growth of dimple population}\label{growth}
\vspace{-0.15cm}
Here we consider the process shown in Fig \ref{processes}(a). Two reservoir atoms with momenta $\vec{p}$ and $\vec{q}$ collide with each other. One of the atoms enters the $\vec{n}$-th state in the dimple, and the other atom recoils with momentum $\vec{k}$. The rate of this process depends on the following factors: (i) It is proportional to the occupation of the momentum states $\vec{p}$ and $\vec{q}$ at the origin, which is given by the Boltzmann factor $\exp[-\beta_r ((p^2 + q^2)/2 m - 2\mu_r)]$. (ii) Due to the quantum-mechanical symmetry of identical Bosons, the likelihood of scattering into the $\vec{n}$-th state is enhanced by a factor of $N_{\vec{n}}$, the number of Bosons already present in the $\vec{n}$-th state. This gives rise to a Bose stimulation factor $1 + N_{\vec{n}}$. (iii) The rate is proportional to $U_0^2$, where $U_0 = 4 \pi \hbar^2 a / m$ is the scattering amplitude for weakly interacting Bosons, $a$ being the s-wave scattering length \cite{pethick}. This factor originates from the overlap of initial and final states in Fermi's golden rule. Since all four single-particle states involved in the collision have definite momenta, the overlap also produces a delta function which imposes conservation of momentum. Finally, we must conserve energy. Combining these factors we can write down the total rate at which atoms enter the $\vec{n}$-th dimple state via such processes:
\begin{align}
\nonumber \bigg(\frac{dN_{\vec{n}}}{dt}\bigg)^g_{in} =& \hspace{0.05cm}\frac{2 \pi}{\hbar} U_0^2 (1 + N_{\vec{n}}) \int^{\prime} \frac{d^3 p d^3 q}{(2 \pi \hbar)^6} \; e^{-\beta_r (\frac{p^2+q^2}{2m} - 2\mu_r)}\\
&\hspace{-1.1cm} \times \delta\bigg(\frac{p^2 + q^2 - \big(\vec{p} + \vec{q} - (2 \pi \hbar/l_d) \vec{n}\big)^2}{2m} - \varepsilon_{n}\bigg). \label{gin}
\end{align}
Here the prime restricts the initial momenta $\vec{p}$ and $\vec{q}$ to regions where $p^2, q^2 < 2 m \varepsilon_t$.

In the reverse process, a reservoir atom collides with an atom in the $\vec{n}$-th dimple state, and both enter the reservoir. The rate at which such processes decrease $N_{\vec{n}}$ can be written using similar reasoning as above:
\begin{align}
\nonumber \bigg(\frac{dN_{\vec{n}}}{dt}\bigg)^g_{out} \hspace{-0.1cm} =& \hspace{0.05cm} -\frac{2 \pi}{\hbar} U_0^2 N_{\vec{n}} \int^{\prime} \frac{d^3 p d^3 q}{(2 \pi \hbar)^6} \; e^{-\beta_r (\frac{p^2+q^2}{2m} - \varepsilon_n - \mu_r)}\\
& \hspace{-1.1cm} \times \delta\bigg(\frac{p^2 + q^2 - \big(\vec{p} + \vec{q} - (2 \pi \hbar/l_d) \vec{n}\big)^2}{2m} - \varepsilon_{n}\bigg). \label{gout}
\end{align}
The net growth rate of $N_{\vec{n}}$ can now be found by summing Eqs. (\ref{gin}) and (\ref{gout}).

In the forward process described by Eq. (\ref{gin}), when the atom recoiling with momentum $\vec{k}$ has energy exceeding $\varepsilon_t$, i.e., $k^2 / 2m = (p^2 + q^2)/2m + \varepsilon_d - E_n > \varepsilon_t$, it is lost from the trap. We call such collisions ``one-way collisions" since they do not have any reverse process. Whereas collisions in which $k^2 / 2m < \varepsilon_t$ can happen both ways. We call such collisions ``two-way collisions." A one-way collision reduces the number of atoms in the reservoir ($N_r$) by 2, whereas a two-way collision reduces $N_r$ by 1. Thus we write
\begin{eqnarray}
\bigg(\frac{dN_{\vec{n}}}{dt}\bigg)^g &=& \bigg(\frac{dN_{\vec{n}}}{dt}\bigg)^g_{1} + \bigg(\frac{dN_{\vec{n}}}{dt}\bigg)^g_{2}\;,\label{ng}\\
\bigg(\frac{dN_r}{dt}\bigg)^g &=& -\sum_{\vec{n}} \bigg[2 \hspace{0.05cm} \bigg(\frac{dN_{\vec{n}}}{dt}\bigg)^g_{1} + \bigg(\frac{dN_{\vec{n}}}{dt}\bigg)^g_{2} \hspace{0.05cm} \bigg],\label{nrg}
\end{eqnarray}
with explicit expressions for these terms in Appendix \ref{growthapp}.

In a two-way collision the reservoir energy $E_r$ increases by $\varepsilon_d - E_n$. This leads to heating. The net rate of increase of $E_r$ due to two-way collisions can be written as
\begin{equation}
\bigg(\frac{dE_r}{dt}\bigg)^g_2 = \sum_{\vec{n}} (\varepsilon_d - E_n)  \bigg(\frac{dN_{\vec{n}}}{dt}\bigg)^g_{2}\;.
\label{Erg2}
\end{equation}
On the other hand, in a one-way collision between two atoms of momenta $\vec{p}$ and $\vec{q}$, their total energy $(p^2 + q^2) / 2m$ is lost from the reservoir. Depending on whether this energy is greater or less than twice the average particle energy in the reservoir, such a collision cools down or heats up the reservoir. We can obtain the rate at which $E_r$ decreases due to one-way collisions which populate the $\vec{n}$-th dimple state by using arguments similar to those preceding Eq. (\ref{gin}):
\begin{align}
\nonumber \hspace{-0.075cm}\bigg(\frac{dE_{r}}{dt}\bigg)^g_{1,\vec{n}} \hspace{-0.05cm}=\hspace{-0.08cm} -\frac{2 \pi}{\hbar} U_0^2 (1 + N_{\vec{n}}) \hspace{-0.07cm} \int^{\prime\prime} \hspace{-0.12cm} \frac{d^3 p d^3 q}{(2 \pi \hbar)^6} \; e^{-\beta_r (\frac{p^2+q^2}{2m} - 2\mu_r)}\\
\times \hspace{0.02cm} \frac{p^2 + q^2}{2m}\hspace{0.1cm}\delta\bigg(\frac{p^2 + q^2 - \big(\vec{p} + \vec{q} - (2 \pi \hbar/l_d) \vec{n}\big)^2}{2m} - \varepsilon_{n}\bigg). \label{Erg1n}
\end{align}
Here the double prime indicates that the initial momenta satisfy $p^2,q^2 < 2 m \varepsilon_t$ and $p^2 + q^2 > 2 m  (\varepsilon_t - \varepsilon_d + E_n)$. Appendix \ref{growthapp} reduces Eq. (\ref{Erg1n}) to a lower dimensional integral. The net rate of change of $E_r$ is given by
\begin{equation}
\bigg(\frac{dE_{r}}{dt}\bigg)^g = \bigg(\frac{dE_{r}}{dt}\bigg)^g_{2} + \sum_{\vec{n}} \bigg(\frac{dE_{r}}{dt}\bigg)^g_{1,\vec{n}}\;.
\label{Erg}
\end{equation}

Due to symmetry, the population of the dimple states depend only on their energy, as can be verified from Eqs. (\ref{ng1}) and (\ref{ng2}). This allows us to describe them by a continuous distribution function in energy $f (\tilde{E},t)$: the number of atoms in the energy interval $d \tilde{E}$ at time $t$ equals $\mathcal{N} f(\tilde{E},t) d \tilde{E}$. Using this definition we can relate $f(\tilde{E},t)$ to $N_{\vec{n}} (t)$ via the density of states $D(\tilde{E})$: $f(\tilde{E},t) = D(\tilde{E}) N_{\vec{n}} (t) / \mathcal{N}$, where $\tilde{E}_n = \tilde{E}$. $D(\tilde{E})$ can be obtained by noting that $\tilde{E}_n = (\pi / \tilde{l}_d^2) n^2$ (see Eq. (\ref{energyntilde})), which yields $D(\tilde{E}) = 2 \tilde{l}_d^3 (\tilde{E} / \pi)^{1/2}$. We can then express Eq. (\ref{ng}) as equations of motion for $f(\tilde{E},t)$. The characteristic time in these equations is $\tau_{\mbox{\scriptsize{coll}}} \Omega$, the product of the collision time and the volume ratio of the reservoir to the dimple.

To account for condensation, we treat $N_{\vec{0}} (t)$ separately from $f(\tilde{E},t)$, and define $f_0 (t) \equiv N_{\vec{0}} (t) / \mathcal{N}$ as the condensate fraction. Eqs. (\ref{gin}) and (\ref{gout}) then give equations of motion for $f_0 (t)$, which can be written in terms of one-way and two-way collisions.

The reservoir fraction $f_r (t)$ defined in Eq. (\ref{frdef}) then evolves according to
\begin{align}
\nonumber \bigg(\frac{df_{r}}{dt}\bigg)^g =&\hspace{0.05cm} -2 \bigg(\frac{df_{0}}{dt}\bigg)^g_{1} - \bigg(\frac{df_{0}}{dt}\bigg)^g_{2} \\
&\hspace{-1cm} - \int_{0}^{\tilde{\varepsilon}_d} d \tilde{E} \bigg[2 \bigg(\frac{\partial f (\tilde{E}, t)}{\partial t}\bigg)^g_{1} + \bigg(\frac{\partial f (\tilde{E}, t)}{\partial t}\bigg)^g_{2}\hspace{0.05cm}\bigg] , \label{frg}
\end{align}
and similar expressions hold for the relative energy in the reservoir $e_r (t)$ defined in Eq. (\ref{er}).

\subsubsection{Redistribution of dimple population}\label{redistribution}

Here we examine two-body collisions of the kind illustrated in Fig \ref{processes}(b), where a reservoir atom of momentum $\vec{p}$ exchanges energy with a dimple atom in state $\vec{n}_1$, sending it to a different state $\vec{n}_2$. The rate of such processes can be calculated using reasoning similar to that outlined at the beginning of section \ref{growth}:
\begin{align}
\nonumber \frac{dN_{\vec{n}_1 \to \vec{n}_2}}{dt} = \frac{2 \pi}{\hbar} U_0^2 N_{\vec{n}_1} (1 + N_{\vec{n}_2}) \frac{1}{l_d^3} \int^{\prime} \hspace{-0.2cm} \frac{d^3 p}{(2 \pi \hbar)^3} \; e^{-\beta_r (\frac{p^2}{2m} - \mu_r)}\\
\times \delta\bigg(\frac{p^2}{2m} + E_{n_1} - \frac{\big(\vec{p} + (2 \pi \hbar / l_d) (\vec{n}_1 - \vec{n}_2)\big)^2}{2m} - E_{n_2}\bigg), \label{n1n2}
\end{align}
where the prime denotes the condition $p^2 < 2 m \varepsilon_t$. The net rate of change of $N_{\vec{n}}$ due to such collisions is then
\begin{equation}
\bigg(\frac{dN_{\vec{n}}}{dt}\bigg)^r = \sum_{\vec{n}^{\prime} \neq \vec{n}} \frac{dN_{\vec{n}^{\prime} \to \vec{n}}}{dt} - \frac{dN_{\vec{n} \to \vec{n}^{\prime}}}{dt}\;. \label{nnr}
\end{equation}
Once again there are one-way and two-way collisions. In Eq. (\ref{n1n2}), if the final energy of the reservoir atom is sufficiently large, $p^2 / 2m + E_{n_1} - E_{n_2} > \varepsilon_t$, it is lost from the trap. Such one-way collisions happen only when the dimple atom is transferred to a much lower energy level. As in our prior sections, collisions in which the above condition is not satisfied happen both ways. These do not change the number of atoms in the reservoir ($N_r$), but can change their average energy, thus changing the reservoir temperature $T_r$. Denoting the rates of one-way and two-way collisions by $R^{(1)}_{\vec{n},\vec{n}^{\prime}}$ and $R^{(2)}_{\vec{n},\vec{n}^{\prime}}$, we write
\begin{eqnarray}
\bigg(\frac{dN_{\vec{n}}}{dt}\bigg)^r &=& \sum_{\vec{n}^{\prime} \neq \vec{n}} R^{(1)}_{\vec{n},\vec{n}^{\prime}} + R^{(2)}_{\vec{n},\vec{n}^{\prime}}\;, \label{nnr12}\\
\bigg(\frac{dN_r}{dt}\bigg)^r &=& -\hspace{-0.1cm}\sum_{\substack{\vec{n}^{\prime}, \vec{n} \\ E_{n^{\prime}} > E_n}} \hspace{-0.1cm} R^{(1)}_{\vec{n},\vec{n}^{\prime}}\;. \label{nrr}
\end{eqnarray}

In a one-way collision, the energy in the reservoir ($E_r$) decreases by an amount $p^2/2m$. Therefore we obtain the net rate of change of $E_r$ due to one-way collisions as
\begin{equation}
\bigg(\frac{dE_r}{dt}\bigg)^r_1 = \sum_{\substack{\vec{n}^{\prime}, \vec{n} \\ E_{n^{\prime}} > E_n}} \bigg(\frac{dE_r}{dt}\bigg)^r_{1,\vec{n}^{\prime} \to \vec{n}} \;,\label{Err1}
\end{equation}
where
\begin{align}
\nonumber \bigg(\frac{dE_r}{dt}\bigg)^r_{1,\vec{n}^{\prime} \to \vec{n}} \hspace{-0.2cm}= -\frac{2 \pi z_r}{\hbar l_d^3} U_0^2 N_{\vec{n}^{\prime}} (1 + N_{\vec{n}}) \int^{\prime\prime} \hspace{-0.15cm} \frac{d^3 p}{(2 \pi \hbar)^3} \; e^{-\beta_r \frac{p^2}{2m}}\\
\times \frac{p^2}{2m}\delta\bigg(\frac{p^2}{2m} + E_{n^{\prime}} - \frac{\big(\vec{p} + (2 \pi \hbar / l_d) (\vec{n}_1 - \vec{n}_2)\big)^2}{2m} - E_{n}\bigg). \label{Errn1n2}
\end{align}
Here the double prime imposes the condition $2 m (\varepsilon_t + E_{n} - E_{n^{\prime}}) < p^2 < 2 m \varepsilon_t$. Whereas, when a two-way collision transfers a dimple atom from state $\vec{n}^{\prime}$ to state $\vec{n}$, the reservoir energy changes by $E_{n^{\prime}} - E_n$. Thus,
\begin{equation}
\bigg(\frac{dE_r}{dt}\bigg)^r_2 = \sum_{\substack{\vec{n}^{\prime}, \vec{n} \\ E_{n^{\prime}} > E_n}} \hspace{-0.1cm} (E_{n^{\prime}} - E_n) \hspace{0.05cm} R^{(2)}_{\vec{n},\vec{n}^{\prime}}\;.\label{Err2}
\end{equation}
The total change in $E_r$ is then found by adding the two contributions (Eqs. (\ref{Err1}) and (\ref{Err2})).

Describing the dimple states in terms of a continuous distribution function $f(\tilde{E}, t)$, as defined in section \ref{growth}, we can express Eq. (\ref{nnr12}) as
\begin{equation}
\bigg(\frac{\partial f(\tilde{E},t)}{\partial t}\bigg)^r = \int_{0}^{\tilde{\varepsilon}_d} d \tilde{E^{\prime}} \big(\mathcal{R}_1 (\tilde{E}, \tilde{E}^{\prime}) + \mathcal{R}_2 (\tilde{E}, \tilde{E}^{\prime})\big) \;,\label{fr}
\end{equation}
where $\mathcal{R}_1 (\tilde{E}, \tilde{E}^{\prime})$ and $\mathcal{R}_2 (\tilde{E}, \tilde{E}^{\prime})$ are given in Appendix \ref{redistributionapp}. As would be expected, we find that the characteristic time for these processes is the collision time $\tau_{\mbox{\scriptsize{coll}}}$. The rate equations for the condensate fraction $f_0$, the reservoir fraction $f_r$, and the relative energy in the reservoir $e_r$ can be obtained likewise from Eqs. (\ref{nnr12})$-$(\ref{Err2}).

To simulate the overall dynamics incorporating both the growth and the redistribution process, we add the corresponding equations of motion describing the two processes. In the next section, we present our numerical results. Three-body processes will be discussed in Sec. \ref{threebody}.

\section{Results}\label{results}

\subsection{Initial dynamics and thermalization}\label{thermalization section}

\begin{figure}[h]
\includegraphics[width=\columnwidth]{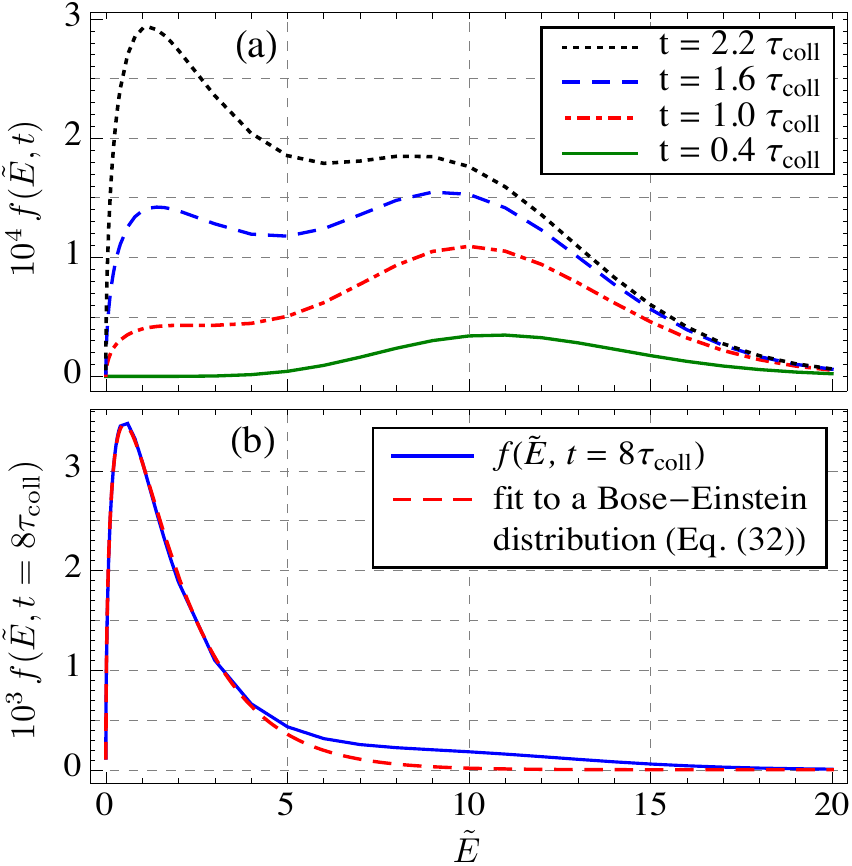}
\caption{(Color online) Time-evolution of the atom distribution in the dimple $f(\tilde{E},t)$ for $0 < t \lesssim 10 \tau_{\mbox{\scriptsize{coll}}}$, where $\tilde{E} \equiv E/ k_B T_{r0}$ is the energy measured from the bottom of the dimple scaled by the initial reservoir temperature. (a) Initial population growth occurs near $\tilde{E} = \tilde{\varepsilon}_d / 2$. These atoms are quickly transferred to lower energy states, giving rise to a hump near $\tilde{E} = 0$ which grows rapidly. Around $t \approx 1.6 \tau_{\mbox{\scriptsize{coll}}}$, $f(\tilde{E},t)$ has a bimodal shape for $\tilde{\varepsilon}_d \gtrsim 8$, which should show up in time-of-flight experiments. (b) $f(\tilde{E},t = 8 \tau_{\mbox{\scriptsize{coll}}})$ is well-fit by a thermal Bose-Einstein distribution given by Eq. (\ref{fit}). Parameter values used for plotting are $\rho_i = 0.05$, $\Omega=2000$, $\tilde{\varepsilon}_d = 20$, $\tilde{\varepsilon}_t = 10$, and $\tilde{l}_d = 100$. At $T_{r0} = 1\hspace{0.05cm} \mu$K, these give $\tau_{\mbox{\scriptsize{coll}}} \approx 11$ ms for ${}^{87}$Rb.}
\label{thermalization}
\end{figure}

Figure \ref{thermalization} shows how the particle distribution in the dimple $f(\tilde{E},t)$ evolves for $0 < t \lesssim 10 \tau_{\mbox{\scriptsize{coll}}}$ after the dimple is turned on at $t=0$. Although the plots correspond to specific initial conditions stated in the caption, we observe the same general features for other choices of parameters. We find that the reservoir particles predominantly scatter into states whose energies are near half the dimple depth, creating a hump in $f(\tilde{E},t)$. Such a highly non-equilibrium distribution does not last for long. Within a few $\tau_{\mbox{\scriptsize{coll}}}$, the processes in Fig. \ref{processes}(b) transfer these atoms to lower energy states near the bottom of the dimple. This generates a hump in $f(\tilde{E},t)$ near $\tilde{E} =0$, which grows and soon overtakes the hump near $\tilde{E} = \tilde{\varepsilon}_d / 2$. As a result, around $t \sim 1.6 \tau_{\mbox{\scriptsize{coll}}}$, $f(\tilde{E},t)$ has a bimodal shape. The peaks are more distinct for larger dimple depths and are hard to resolve when $\tilde{\varepsilon}_d \lesssim 8$. This non-equilibrium stage lasts for a few collision times. For ${}^{87}$Rb with $\rho_i = 0.05$ and $T_{r0} = 1\hspace{0.05cm} \mu$K, the collision time is $\tau_{\mbox{\scriptsize{coll}}} \approx 11$ ms, sufficiently long that one can experimentally resolve these dynamics. In time-of-flight imaging, the bimodal shape of $f(\tilde{E}, t)$ should produce two expanding shells of atoms.

For $t \gtrsim 8 \tau_{\mbox{\scriptsize{coll}}}$, we find that $f(\tilde{E},t)$ is well-approximated by a thermal distribution. This is seen in Fig. \ref{thermalization}(b) where we fit $f(\tilde{E},t = 8 \tau_{\mbox{\scriptsize{coll}}})$ to a Bose-Einstein distribution truncated at $\tilde{E} = \tilde{\varepsilon}_d$:
\begin{equation}
f(\tilde{E}) = \frac{D(\tilde{E}) / \mathcal{N}}{e^{\beta_d (-\varepsilon_d + E - \mu_d)} - 1} = \frac{1}{\rho_i \Omega} \frac{2 (\tilde{E} / \pi)^{1/2}}{e^{\tilde{\beta}_d (-\tilde{\varepsilon}_d + \tilde{E} - \tilde{\mu}_d)} - 1}\;.
\label{fit}
\end{equation}
Here $\beta_d \equiv \tilde{\beta}_d \beta_{r0} = 1/k_B T_d$ and $\mu_d \equiv \tilde{\mu}_d / \beta_{r0}$ denote the inverse temperature and chemical potential of the dimple. The high-energy tail of $f(\tilde{E},t)$ takes a little longer to thermalize. Once the density in the dimple exceeds that of the reservoir, the timescales for redistribution become much shorter than those for growth. Thus we find quasi-thermal equilibrium inside the dimple for $t \gtrsim 8 \tau_{\mbox{\scriptsize{coll}}}$, though $\beta_d$ and $\mu_d$ change with time.

\begin{figure}[t]
\includegraphics[width=\columnwidth]{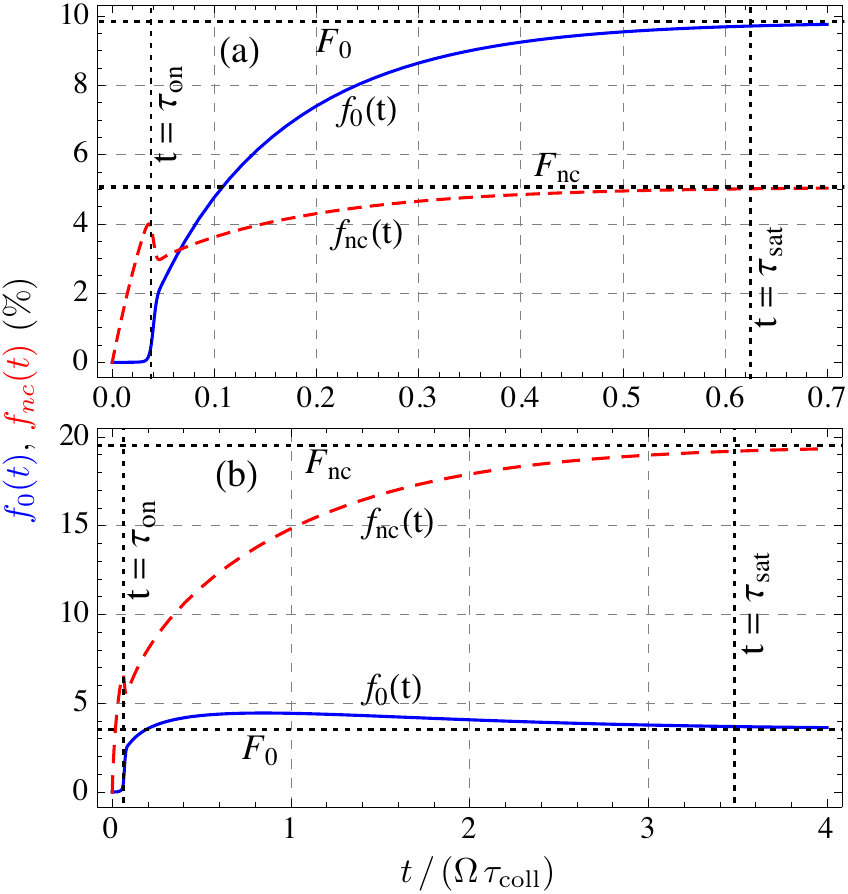}
\caption{(Color online) Evolution of the condensate fraction $f_0$ (solid blue) and the non-condensate fraction $f_{nc}$ (dashed red) in the dimple for $\rho_i = 0.05$, $\Omega=2000$, and $\tilde{l}_d = 20$, assuming no three-body loss. $f_0 (t)$ grows slowly at first until Bose stimulation takes over at the onset time $\tau_{\mbox{\scriptsize{on}}}$. The populations reach equilibrium at a much later time $\tau_{\mbox{\scriptsize{sat}}}$. $F_0$ and $F_{nc}$ denote the saturation values of $f_0$ and $f_{nc}$ respectively. (a) $\tilde{\varepsilon}_d = 7$: when the dimple depth is small, $f_0 (t)$ monotonically increases toward $F_0$. (b) $\tilde{\varepsilon}_d = 28$: at large dimple depths, $f_0 (t)$ overshoots $F_0$ after $t = \tau_{\mbox{\scriptsize{on}}}$ before coming down again.}
\label{fvst}
\vspace{-0.15cm}
\end{figure}

We find that the condensate fraction $f_0$ grows very slowly at first until it becomes large enough that Bose stimulation can take over. This gives rise to a noticeable time delay in the onset of condensation, marked as $\tau_{\mbox{\scriptsize{on}}}$ in Fig. \ref{fvst} where we plot $f_0 (t)$ for two different sets of parameter values.
Figure \ref{fvst} also shows the evolution of the total non-condensate fraction in the dimple $f_{nc}(t) = \int_0^{\tilde{\varepsilon}_d} d \tilde{E} f (\tilde{E},t)$. After $t = \tau_{\mbox{\scriptsize{on}}}$, $f_0 (t)$ grows rapidly due to Bose enhancement. Part of this enhanced growth comes from atoms in low-lying excited states scattering to the ground state via two-body collisions with reservoir atoms. This redistribution causes a sudden dip in $f_{nc}(t)$ just after $t = \tau_{\mbox{\scriptsize{on}}}$. In the absence of three-body loss, $f_0 (t)$ and $f_{nc} (t)$ reach their respective saturation values $F_0$ and $F_{nc}$ at a much later time $t = \tau_{\mbox{\scriptsize{sat}}}$. When $\tilde{\varepsilon}_d$ is small, $f_0 (t)$ monotonically approaches $F_0$ from below (Fig. \ref{fvst}(a)), whereas for large dimple depths, $f_0 (t)$ overshoots $F_0$ shortly after $t = \tau_{\mbox{\scriptsize{on}}}$ (Fig. \ref{fvst}(b)). In both cases, however, the reservoir fraction $f_r (t)$ monotonically decreases from 1 toward its saturation value $F_r$.
\subsection{Results for infinite trap depth}\label{infinite trap depth}

When the trap is sufficiently deep, no particles can be lost. Therefore only two-way collisions are present and we have the relation $f_r (t) + f_0 (t) + f_{nc} (t) = 1$. We can simplify the calculation further by noting our previous observation that the dimple thermalizes very quickly compared to the growth rate of its atom population. Thus we assume that $f(\tilde{E},t)$ is always described by a thermal distribution as given in Eq. (\ref{fit}), where the temperature ($T_d$) and the chemical  potential ($\mu_d$) of the dimple vary with time. This approximation allows us to rapidly simulate the dynamics for a wide range of parameter values and reproduces all features of the full model for $t \gtrsim 10 \tau_{\mbox{\scriptsize{coll}}}$.
\vspace{-0.3cm}
\subsubsection{Long time behavior}\label{long time behavior}
\begin{figure}[t]
\includegraphics[width=\columnwidth]{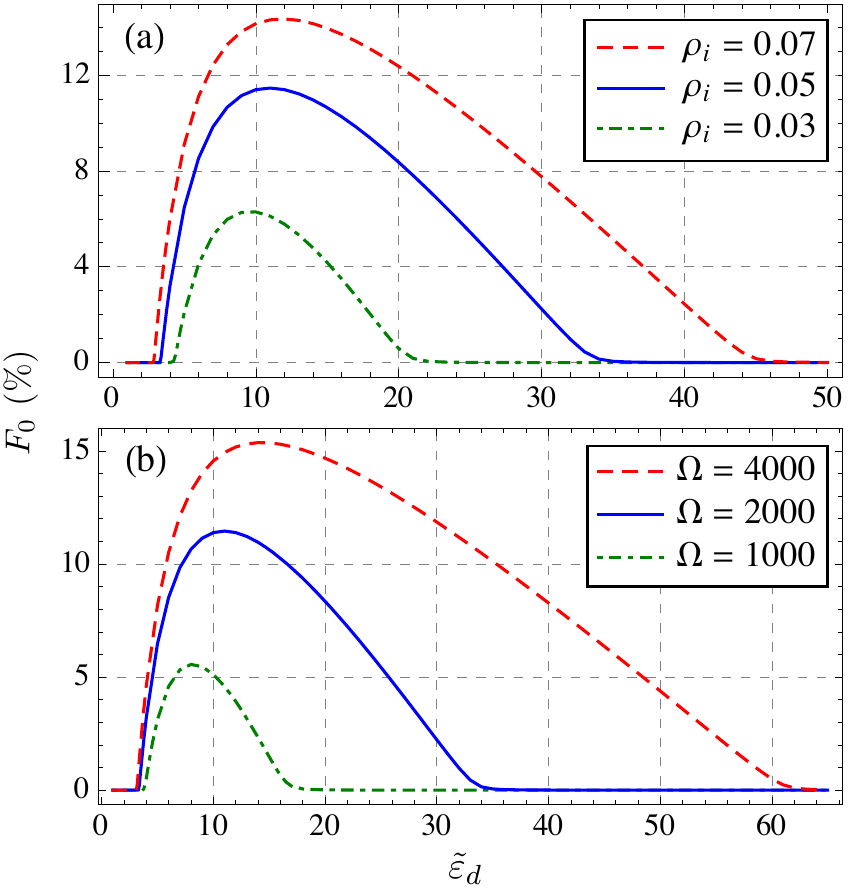}
\caption{(Color online) Variation of the final condensate fraction $F_0$ with the dimple depth $\tilde{\varepsilon}_d \equiv \varepsilon_d / k_B T_{r0}$ for different choices of the initial phase space density $\rho_i$ and the volume ratio $\Omega$ when the trap depth is infinite and there is no three-body loss. In (a), $\Omega = 2000$ and in (b), $\rho_i = 0.05$. When $\tilde{\varepsilon}_d < |\tilde{\mu}_{r0}| = |\ln (\rho_i)|$, the atoms do not condense. As $\tilde{\varepsilon}_d$ is increased beyond this threshold, the atom density in the dimple grows, producing a larger $F_0$. However, atoms scattering into a deeper dimple also cause more heating, which prevents condensation at large $\tilde{\varepsilon}_d$. Consequently, there exists an optimal dimple depth $\tilde{\varepsilon}_d^*$ which yields the maximum condensate fraction. Larger $\rho_i$ and $\Omega$ increase the atom density in the dimple without causing much change in the final temperature, hence give a larger $F_0$. The optimal dimple depth $\tilde{\varepsilon}_d^*$ also grows with both $\rho_i$ and $\Omega$.}
\label{F0}
\end{figure}

\vspace{-0.1cm}
Figure \ref{F0} shows the variation of $F_0$ with $\tilde{\varepsilon}_d$ for different values of the initial phase space density $\rho_i$ and the volume ratio $\Omega$. The different features in the plots can be explained by the following model: as particles are scattered into the negative energy states of the dimple, those remaining in the reservoir have a higher total energy. Therefore the temperature $T_r$ increases and the chemical potential $\mu_r$ drops (see Eqs. (\ref{number2}) and (\ref{energy})). In equilibrium, $T_r = T_d \equiv T_f$ and $\mu_r = \mu_d \equiv \mu_f \leq -\varepsilon_d$, since we are considering Bosonic atoms. At a given temperature $T_f$, there is an upper limit to the number of non-condensate particles the dimple can hold, which occurs when $\mu_f = -\varepsilon_d$:
\begin{equation}
f_{nc}^{max} \approx \frac{1}{\rho_i \Omega}\int_0^{\infty} d \tilde{E} \;\frac{2 (\tilde{E} / \pi)^{1/2}}{e^{\tilde{\beta}_f \tilde{E}} - 1} = \frac{\zeta(3/2)}{\rho_i \Omega \tilde{\beta}_f^{3/2}}\hspace{0.05cm},
\label{fncmax}
\end{equation}
where $\beta_f \equiv 1/k_B T_f$. Consequently, only if $1 - F_r \geq f_{nc}^{max}$, will we get condensation. The condensate fraction will be $F_0 = 1 - F_r - f_{nc}^{max}$ and the chemical potential will be $\mu_f = -\varepsilon_d$.
\begin{figure}[t]
\includegraphics[width=\columnwidth]{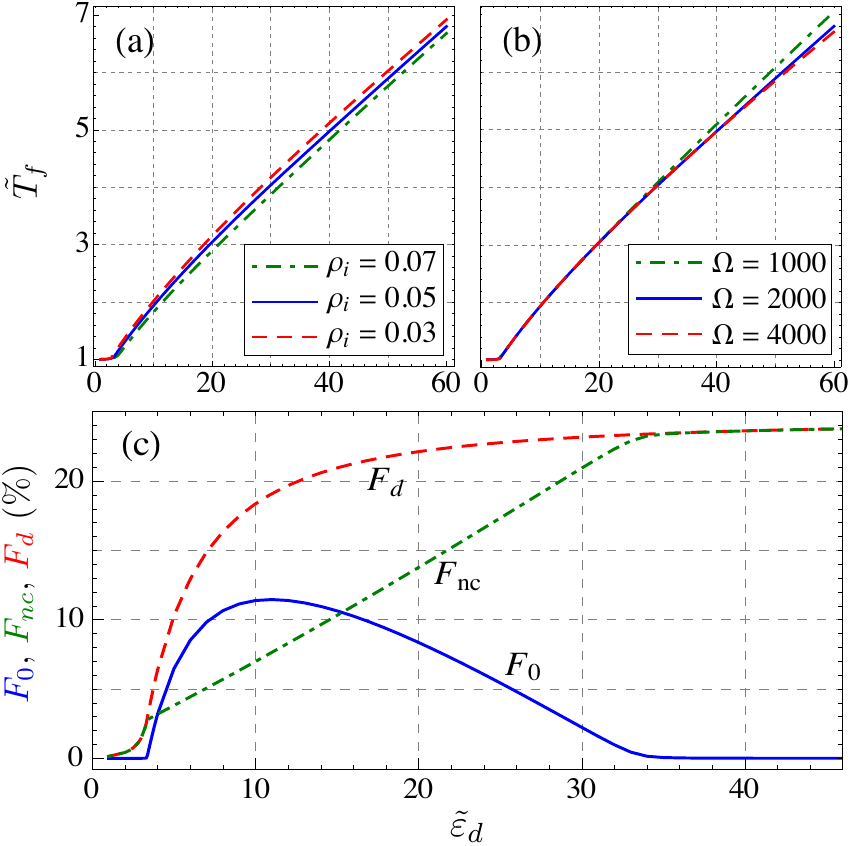}
\caption{(Color online) (a) and (b) Variation of the final temperature $\tilde{T}_f \equiv T_f / T_{r0}$ with the dimple depth $\tilde{\varepsilon}_d \equiv \varepsilon_d / k_B T_{r0}$ for different choices of $\rho_i$ and $\Omega$. In (a) $\Omega = 2000$, and in (b), $\rho_i = 0.05$. When $\tilde{\varepsilon}_d < |\tilde{\mu}_{r0}|$, very few atoms scatter into the dimple and $T_f \approx T_{r0}$. As $\tilde{\varepsilon}_d$ is increased, atoms scattering into the dimple cause more heating of the reservoir. Thus $\tilde{T}_f$ increases monotonically with $\tilde{\varepsilon}_d$. (c) Variation of the final condensate fraction $F_0$ (solid blue), the final non-condensate fraction $F_{nc}$ (dot-dashed green), and their sum $F_d$ (dashed red) with the dimple depth $\tilde{\varepsilon}_d$ for $\rho_i = 0.05$ and $\Omega = 2000$. When $\mu_{r0} < -\varepsilon_d$, no condensation occurs, and $F_0 = 0$. At larger dimple depths, the phase space density in the dimple becomes large enough to reach condensation. In this regime, the final chemical potential lies at the bottom of the dimple, $\mu_f \approx -\varepsilon_d$, and the non-condensate fraction in the dimple follows the standard expression for a Bose gas, $F_{nc} = \zeta(3/2) / \rho_i \Omega \tilde{\beta}_f^{3/2} \propto \tilde{\varepsilon}_d$. Deeper dimples give a larger $\tilde{T}_f$, causing $F_{nc}$ to grow monotonically. When $\tilde{\varepsilon}_d$ becomes very large, excessive heating prevents condensation. Thereafter $\mu_f$ decreases below $-\varepsilon_d$, causing $F_{nc}$ to saturate.}  
\label{Tf}
\end{figure}
The chemical potential of the reservoir must monotonically decrease as particles scatter into the dimple. Thus we only find condensation if $\ln(\rho_i) = \tilde{\mu}_{r0} > -\tilde{\varepsilon}_d$. This behavior is illustrated in Fig. \ref{F0}. As $\tilde{\varepsilon}_d$ increases from this threshold, the condensate fraction grows. This growth occurs because the lower final chemical potential implies a lower density of the reservoir atoms (and hence a larger number of atoms in the dimple). However, deeper dimples also lead to more heating of the reservoir. If $\tilde{\varepsilon}_d$ is too large, this heating prevents condensation. Thus $F_0$ is non-monotonic, and there exists an optimal dimple depth, $\tilde{\varepsilon}_d^*$, for which $F_0$ is maximum. The $F_0$ vs $\tilde{\varepsilon}_d$ curves for different $\rho_i$ and $\Omega$ can be well-reproduced by assuming $\mu_f = -\varepsilon_d$ and imposing conservation of energy and particle number. When $k_B T_f$ is small relative to $\varepsilon_d$, this yields
\begin{equation}
F_r \tilde{T}_f - \frac{\tilde{\varepsilon}_d}{3} (1 - F_r) + f_{nc}^{max}\hspace{0.05cm} \tilde{T}_f \hspace{0.03cm}\frac{\zeta(5/2)}{2 \hspace{0.05cm} \zeta(3/2)} = 1 \hspace{0.05cm},
\label{F0vsedexplanation}
\end{equation}
with $F_r = e^{-\tilde{\beta}_f \tilde{\varepsilon}_d} / \rho_i \tilde{\beta}_f^3$ and $f_{nc}^{max}$ given by Eq. (\ref{fncmax}). Solving Eq. (\ref{F0vsedexplanation}) for $\tilde{T}_f$ one finds a very weak dependence on $\rho_i$ and virtually no dependence on $\Omega$ for sufficiently large $\Omega$. Therefore, choosing a higher volume ratio does not change the reservoir fraction $F_r$ but decreases the maximum fraction of non-condensate particles in the dimple $f_{nc}^{max}$. In other words, increasing the ratio of the reservoir size to the dimple size increases the local atom density without altering the final temperature, thus producing a larger condensate fraction. Similarly, a large $\rho_i$ decreases both $F_r$ and $f_{nc}^{max}$, hence increasing $F_0$, as one would expect intuitively. This explains why the $F_0$ vs $\tilde{\varepsilon}_d$ curves in Fig. \ref{F0} are higher for greater values of $\Omega$ and $\rho_i$. We also find that the optimal dimple depth $\tilde{\varepsilon}_d^*$ increases with both $\Omega$ and $\rho_i$.

Figures \ref{Tf}(a) and \ref{Tf}(b) show that the final temperature rises linearly with the dimple depth when $\tilde{\varepsilon}_d$ is large. We can understand this behavior by considering the limit of very large $\tilde{\varepsilon}_d$, where $F_0$ is vanishingly small, and the dimple population can be treated classically. Using Eqs. (\ref{frdef}) and (\ref{fit}), one can write $F_r = z \hspace{0.02cm} e^{-\tilde{\beta}_f \tilde{\varepsilon}_d} / \rho_i \tilde{\beta}_f^3$ and $F_{nc} \approx z /\rho_i \Omega \tilde{\beta}_f^3$, with $z = \exp{(\beta_f (\varepsilon_d + \mu_f))}$. Conservation of both energy and particle number then yields
\begin{equation}
F_r = \frac{\tilde{\beta}_f \tilde{\varepsilon}_d + 3 \tilde{\beta}_f - 3/2}{\tilde{\beta}_f \tilde{\varepsilon}_d + 3/2} = \frac{1}{1 + \tilde{\beta}_f^{3/2} e^{\tilde{\beta}_f \tilde{\varepsilon}_d} / \Omega} \hspace{0.05cm}.
\label{Tfvsedexplanation}
\end{equation}
To approximate the solution to this transcendental equation, we replace the right-hand side with a step function. The value of $\tilde{\beta}_f$ at the center of the step (and hence our approximate solution to Eq. (\ref{Tfvsedexplanation})) is found by setting $1/(1 + \tilde{\beta}_f^{3/2} e^{\tilde{\beta}_f \tilde{\varepsilon}_d} / \Omega) = 1/2$. This gives $\tilde{\beta}_f = 1.5 \hspace{0.05cm} \tilde{\varepsilon}_d^{-1} \hspace{0.05cm} W\big((2^{1/3}/ 3)\hspace{0.05cm} \tilde{\varepsilon}_d \hspace{0.05cm}\Omega^{2/3}\big)$, where $W$ denotes the Lambert W function. Since $W$ increases only logarithmically for large arguments, $\tilde{T}_f$ rises linearly with $\tilde{\varepsilon}_d$ when $\tilde{\varepsilon}_d$ is large. Substituting this result into the left-hand size of Eq. (\ref{Tfvsedexplanation}), we see that for large $\tilde{\varepsilon}_d$, $F_r \sim A + B / \tilde{\varepsilon}_d$, where $A$ and $B$ depend logarithmically on $\tilde{\varepsilon}_d$. This structure is apparent in Fig. \ref{Tf}(c) as a saturation of $F_{nc}$. At smaller dimple depths, $F_{nc}$ equals $f_{nc}^{max}$ which grows with $\tilde{\varepsilon}_d$ as $\tilde{\beta}_f$ decreases.

\subsubsection{Timescales}\label{timescales}

As seen from Fig. \ref{fvst}, the condensation dynamics are well-characterized by two timescales: the time needed for the populations to saturate, $\tau_{\mbox{\scriptsize{sat}}}$, and the time which marks the onset of condensation, $\tau_{\mbox{\scriptsize{on}}}$.

\begin{figure}[t]
\includegraphics[width=\columnwidth]{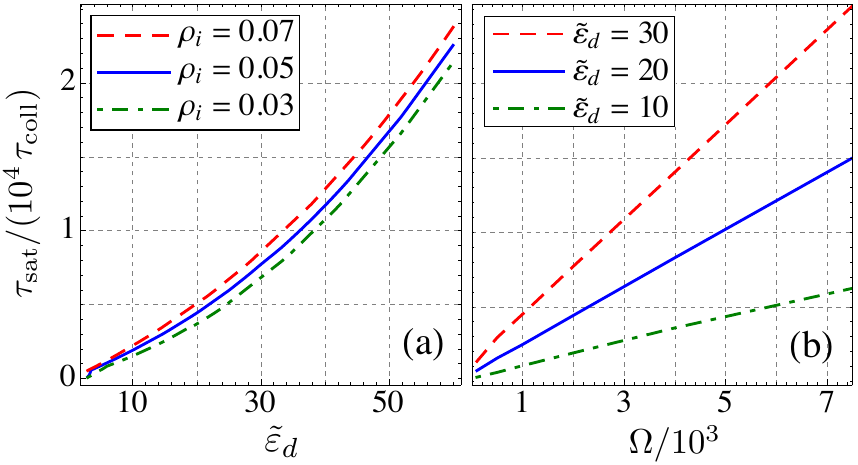}
\caption{(Color online) (a) Variation of the saturation time $\tau_{\mbox{\scriptsize{sat}}}$ with the dimple depth $\tilde{\varepsilon}_d$ for different values of $\rho_i$ when $\Omega = 2000$. The rate of particle scattering into the dimple falls with increasing $\tilde{\varepsilon}_d$ due of smaller overlap between the initial and final states. This results in a larger $\tau_{\mbox{\scriptsize{sat}}}$. Changing the initial phase space density $\rho_i$ alters the collision time $\tau_{\mbox{\scriptsize{coll}}}$, however $\tau_{\mbox{\scriptsize{sat}}} / \tau_{\mbox{\scriptsize{coll}}}$ remains essentially unchanged. This is expected since the dimple is populated via two-body collisions. (b) Variation of $\tau_{\mbox{\scriptsize{sat}}}$ with the volume ratio $\Omega$ at different dimple depths when $\rho_i = 0.05$. $\tau_{\mbox{\scriptsize{sat}}}$ grows almost linearly with $\Omega$ because a larger $\Omega$ increases the total particle number without changing the scattering rate into the dimple. Thus $\Omega \hspace{0.04cm} \tau_{\mbox{\scriptsize{coll}}}$ sets the typical timescale for saturation.}
\label{tausat}
\end{figure}

Chemical equilibrium is reached when the reservoir chemical potential $\mu_r$ crosses below the dimple bottom and merges with the dimple chemical potential $\mu_d$. This is accompanied by the reservoir fraction $f_r (t)$ approaching $F_r$ from above. For concreteness, we define the saturation time $\tau_{\mbox{\scriptsize{sat}}}$ as the time required for $f_r (t)$ to equal $1.002 \hspace{0.03cm} F_r$. In Fig. \ref{tausat}(a) we plot $\tau_{\mbox{\scriptsize{sat}}}$ vs $\tilde{\varepsilon}_d$ for different choices of $\rho_i$. The collapse of the curves for different $\rho_i$ indicates that $\tau_{\mbox{\scriptsize{sat}}}$ is proportional to $\tau_{\mbox{\scriptsize{coll}}}$. This is expected since the dimple is populated via two-body collisions. Additionally, the rate of these two-body collisions decreases as the dimple is made deeper due to reduced overlap between the initial and final states, causing $\tau_{\mbox{\scriptsize{sat}}}$ to rise monotonically. Fig \ref{tausat}(b) shows the variation of $\tau_{\mbox{\scriptsize{sat}}}$ with the volume ratio $\Omega$ for different values of the dimple depth. As evident from the plots, $\Omega \hspace{0.04cm} \tau_{\mbox{\scriptsize{coll}}}$ sets the typical timescale for saturation. This is because a larger $\Omega$ increases the total number of atoms $\mathcal{N}$ without changing the particle scattering rate into the dimple. Thus it takes longer for the dimple population to reach a given fraction of $\mathcal{N}$.

\begin{figure}[t]
\includegraphics[width=\columnwidth]{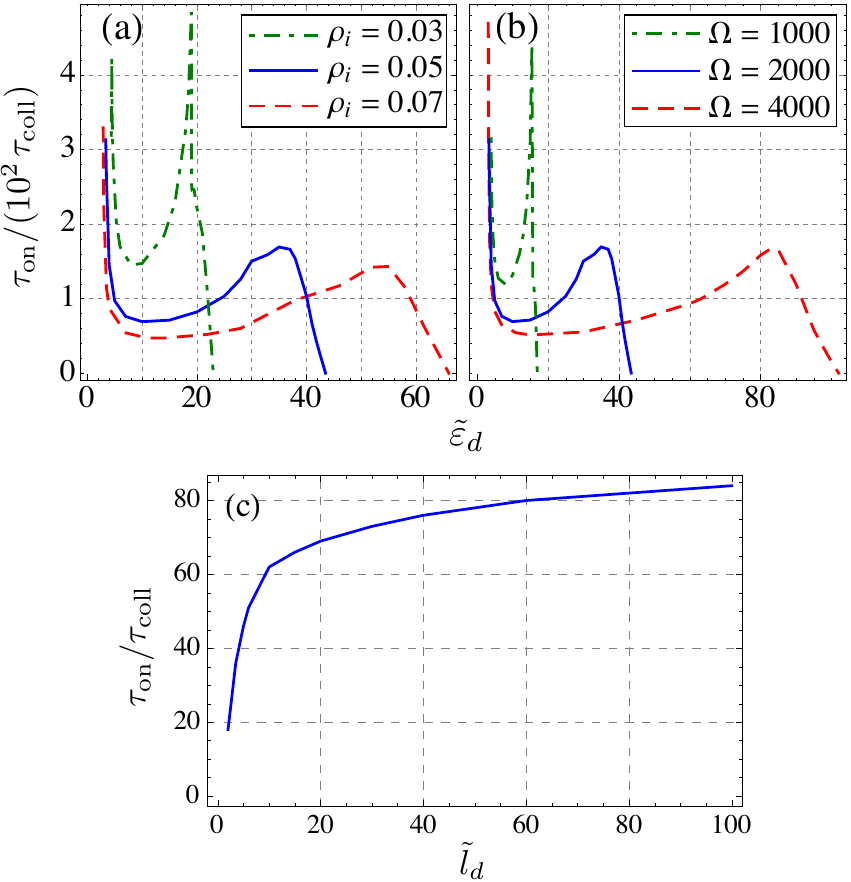}
\caption{(Color online) (a) and (b) Variation of the onset time $\tau_{\mbox{\scriptsize{on}}}$ with the dimple depth $\tilde{\varepsilon}_d$ for different choices of $\rho_i$ and $\Omega$ when the dimple size $\tilde{l}_d \equiv l_d / \lambda_{r0} = 20$, where $\lambda_{r0}$ is the initial thermal wavelength. In (a), $\Omega = 2000$, and in (b), $\rho_i = 0.05$. Within the range of $\tilde{\varepsilon}_d$ where the saturation condensate fraction $F_0$ is significant, $\tau_{\mbox{\scriptsize{on}}}$ is minimum near the optimal dimple depth $\tilde{\varepsilon}_d^*$ which maximizes $F_0$ (see Fig. \ref{F0}). As $\tilde{\varepsilon}_d$ is lowered below $\tilde{\varepsilon}_d^*$, $\tau_{\mbox{\scriptsize{on}}}$ grows and $\tau_{\mbox{\scriptsize{sat}}}$ falls until they become equal at $\tilde{\varepsilon}_d = \tilde{\varepsilon}_d^{>} \approx  -\tilde{\mu}_{r0}$. For smaller dimple depths, condensation does not occur and $F_0 = 0$. $F_0$ also becomes vanishingly small when $\tilde{\varepsilon}_d$ exceeds a large value $\tilde{\varepsilon}_d^{<}$. However, for a range of $\tilde{\varepsilon}_d > \tilde{\varepsilon}_d^{<}$, the condensate fraction rises to significant values before falling to 0. Thus $\tau_{\mbox{\scriptsize{on}}}$ is well-defined in this range. (c) Variation of $\tau_{\mbox{\scriptsize{on}}}$ with $\tilde{l}_d$ for $\rho_i = 0.05$, $\Omega = 2000$, and $\tilde{\varepsilon}_d = 10$. Since the initial growth rate of the condensate fraction is proportional to $1/\tilde{l}_d^3$, $\tau_{\mbox{\scriptsize{on}}}$ increases with $\tilde{l}_d$, saturating when $\tilde{l}_d$ becomes sufficiently large.}
\label{tauon}
\end{figure}

For suitable initial conditions, $\mu_d$ quickly reaches the bottom of the dimple, signaling the onset of condensation. Thereafter $f_0 (t)$ grows rapidly due to Bose stimulation. We define the onset time $\tau_{\mbox{\scriptsize{on}}}$ as the time when the growth rate of $f_0$ increases at the maximum rate, i.e., $d^3 f_0 / d t^3 |_{t = \tau_{\mbox{\tiny{on}}}} = 0$ (see Fig. \ref{fvst}). Figure \ref{tauon} shows how $\tau_{\mbox{\scriptsize{on}}}$ varies with $\tilde{\varepsilon}_d$, $\rho_i$, $\Omega$, and $\tilde{l}_d$. For very small or very large dimple depths, we do not find any condensation, so $f_0 (t)$ is never macroscopic, and it is not sensible to quote an onset time. Interestingly, there are ranges of large $\tilde{\varepsilon}_d$ where $F_0 \approx 0$, but $f_0 (t)$ rises to significant values before falling to 0. Thus $\tau_{\mbox{\scriptsize{on}}}$ is well-defined even for some parameters where $F_0$ is vanishingly small. In the window of $\tilde{\varepsilon}_d$ where $F_0$ is significant, we find that $\tau_{\mbox{\scriptsize{on}}}$ is minimum near the optimal dimple depth $\tilde{\varepsilon}_d^*$. Therefore, choosing $\tilde{\varepsilon}_d = \tilde{\varepsilon}_d^*$ both minimizes the onset time and yields the maximum condensate fraction. As $\tilde{\varepsilon}_d$ is decreased below $\tilde{\varepsilon}_d^*$, $f_0 (t)$ takes longer and longer to take off, but saturates more quickly. Thus $\tau_{\mbox{\scriptsize{on}}}$ rises while $\tau_{\mbox{\scriptsize{sat}}}$ diminishes, until at a particular dimple depth $\tilde{\varepsilon}_d^{>}$, the two timescales become equal. For lower $\tilde{\varepsilon}_d$, the atoms do not condense. This phenomenon shows up in the $\tau_{\mbox{\scriptsize{on}}}$ vs $\tilde{\varepsilon}_d$ curve as an apparent singularity at $\tilde{\varepsilon}_d = \tilde{\varepsilon}_d^{>}$. Increasing either $\rho_i$ or $\Omega$ favors condensation and lowers $\tau_{\mbox{\scriptsize{on}}}$.

The initial growth rate of $f_0$ is proportional to the inverse volume of the dimple measured in units of the thermal wavelength, $1/\tilde{l}_d^3$ (see Eqs. (\ref{f0g1}) and (\ref{f0g2})). Therefore, a higher $\tilde{l}_d$ leads to a smaller onset time. This trend is seen in Fig. \ref{tauon}(c) which shows how $\tau_{\mbox{\scriptsize{on}}}$ varies with $\tilde{l}_d$ when $\rho_i$, $\Omega$, and $\tilde{\varepsilon}_d$ are held fixed. $\tilde{l}_d$ does not affect other features of the dynamics. For large $\tilde{l}_d$, where most experiments operate, $\tau_{\mbox{\scriptsize{on}}}$ also becomes independent of $\tilde{l}_d$. When $\tilde{l}_d \hspace{0.03cm} \tilde{\varepsilon}_d^{1/2} \lesssim 1$, the continuum approximation for the dimple states is not expected to hold (see Eq. (\ref{energyntilde})), and the kinetics should be modeled via a discrete spectrum.

\begin{figure}[t]
\includegraphics[width=\columnwidth]{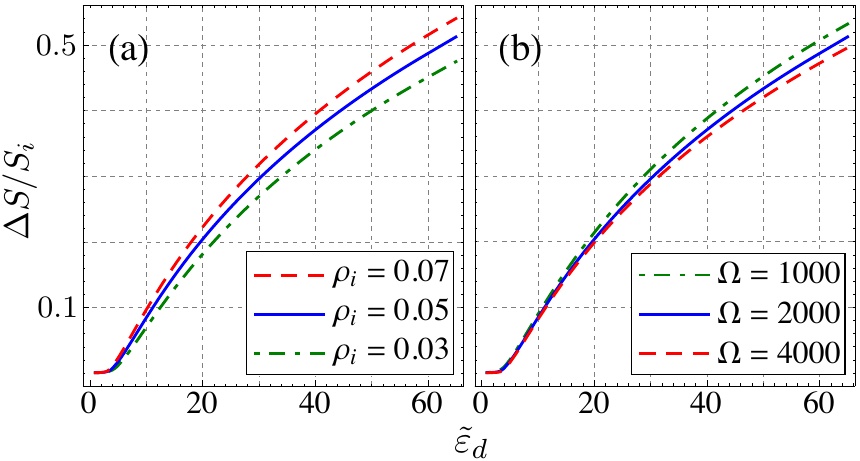}
\caption{(Color online) Relative increase of the total entropy of the reservoir and the dimple, $\Delta S / S_i$, as a function of the dimple depth $\tilde{\varepsilon}_d$ for different values of $\rho_i$ and $\Omega$. In (a), $\Omega = 2000$, and in (b), $\rho_i = 0.05$. The entropy grows because the dimple is turned on suddenly, leading to non-adiabatic dynamics. The dimple loading is more non-adiabatic when $\tilde{\varepsilon}_d$ is large, causing a higher entropy gain. $\Delta S / S_i$ also grows as $\rho_i$ is increased or $\Omega$ is decreased, although this dependence is weak.}
\label{dels}
\end{figure}

We have considered a sudden turn-on of the dimple. Hence the dimple loading is non-adiabatic. A measure of the non-adiabaticity is given by the percentage increase in the total entropy, which we plot in Fig. \ref{dels}. As expected, we find that the dynamics are more non-adiabatic for deeper dimples, with the entropy increasing by $50 \%$ when $\tilde{\varepsilon}_d \approx 60$. The entropy gain also shows a weak dependence on $\rho_i$ and $\Omega$, increasing slowly as $\rho_i$ is increased or $\Omega$ is decreased.

\subsection{Effect of three-body loss}\label{threebody}

Here we incorporate three-body loss into our kinetics model. The loss introduces an additional timescale to the dynamics which depends on the particular atomic species. We will consider the case of ${}^{87}$Rb.

At low temperatures the rate of three-body recombinations is, to a good approximation, proportional to the probability of finding three particles at the same point \cite{pethick}. Thus the loss rate of the total atom density $n(\vec{r})$ is
\begin{equation}
\bigg(\frac{dn(\vec{r})}{dt}\bigg)^l = -L \langle \big( \hat{\Psi}^{\dagger} (\vec{r}) \big)^3 \big( \hat{\Psi} (\vec{r}) \big)^3 \rangle \hspace{0.05cm},
\label{densitylossrate}
\end{equation}
where $L$ denotes the loss coefficient which was measured experimentally for ${}^{87}$Rb as $L = 1.8 \times 10^{-29}$ $\mbox{cm}^6 \hspace{0.05cm} \mbox{s}^{-1}$ \cite{dalibard99}. We write $\hat{\Psi} (\vec{r})$ as the sum of a condensate mean field $\psi_0 (\vec{r})$ and a field $\hat{\psi}_{th} (\vec{r})$ representing thermal fluctuations. Substituting this decomposition into Eq. (\ref{densitylossrate}) and using Wick's theorem to expand, we find \cite{davis11, pethick, kagan85}
\begin{align}
\nonumber \bigg(\frac{dn(\vec{r})}{dt}\bigg)^l = -L \Big[& n_0^3 (\vec{r}) + 9 \hspace{0.05cm} n_0^2 (\vec{r}) \hspace{0.05cm} n_{ex} (\vec{r}) \\
&+ 18 \hspace{0.05cm} n_0 (\vec{r}) \hspace{0.05cm} n_{ex}^2 (\vec{r}) + 6 n_{ex}^3 (\vec{r}) \Big] \hspace{0.05cm},
\label{densitylossrate2}
\end{align}
where $n_0 (\vec{r}) = |\psi_0 (\vec{r}) |^2$ and $n_{ex} (\vec{r}) = \langle \hat{\psi}_{th}^{\dagger} (\vec{r}) \hat{\psi}_{th} (\vec{r}) \rangle$ denote the densities of the condensate and excited-state atoms respectively. $n_{ex} (\vec{r})$ is further decomposed into $n_{nc} (\vec{r})$, the density of the non-condensate atoms in the dimple, and $n_{r} (\vec{r})$, the density of reservoir atoms from Eq. (\ref{nr}). We then derive the decay rates of the individual densities and how these decays contribute to the kinetics (see Appendix \ref{threebodyapp}). In particular, we find that the condensate fraction $f_0 (t)$ evolves as
\begin{equation}
\bigg(\frac{df_0 (t)}{dt}\bigg)^l = -L \frac{\rho_i^2 \Omega^2}{\lambda_{r0}^6} f_0 \Big[f_0^2 + 6 f_0 f^{\prime} + 6 f^{\prime 2} \Big],\label{f0l}
\end{equation}
where $f^{\prime} \equiv f_{nc} + (\tilde{z}_r / \Omega \tilde{\beta}_r^{3/2}) \big(\gamma(3/2,\tilde{\beta}_r \tilde{\varepsilon}_t) / \gamma(3/2, \tilde{\varepsilon}_t)\big)$ and $\gamma$ denotes the lower incomplete gamma function.

\subsubsection*{Results for the case $\varepsilon_t \gg k_B T_{r0}, \varepsilon_d$}

As with the discussion in Sec. \ref{infinite trap depth}, we consider the limit of infinite trap depth $\varepsilon_t \gg k_B T_{r0}, \varepsilon_d$. Here we have loss, and the total number of atoms in the system decays monotonically toward zero. The decay rate depends explicitly on the particle density. The more general case of finite trap depth will be discussed in Sec. \ref{finite trap depth}.

\begin{figure}[t]
\includegraphics[width=\columnwidth]{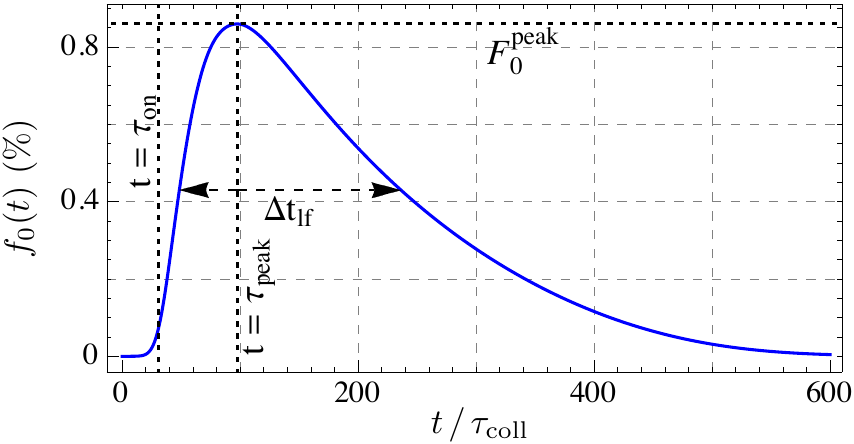}
\caption{Time-evolution of the condensate fraction $f_0$ in the presence of three-body loss for $\tilde{\varepsilon}_d = 8$, $\Omega = 2000$, $\rho_i = 0.2$, and $T_{r0} = 0.1$ $\mu$K. $f_0$ takes off at $t = \tau_{\mbox{\scriptsize{on}}}$ due to Bose stimulation. As the dimple population grows, the atom density in the dimple increases, which leads to a higher three-body recombination rate. At $t = \tau_{\mbox{\scriptsize{peak}}}$, the three-body decay rate of $f_0$ balances the two-body scattering rate into the condensate, and $f_0$ reaches its peak value $F_0^{\mbox{\scriptsize{peak}}}$. As more particles scatter into the dimple, the temperature continues to increase. Three-body losses further heat the system by ejecting more low-energy particles. This heating, combined with the particle loss, causes the chemical potential to drop which decreases $f_0$. Thus we get a finite condensate lifetime $\Delta t_{\mbox{\scriptsize{lf}}}$.}
\label{f0loss}
\end{figure}

As before, we find that the condensate fraction $f_0 (t)$ grows slowly at first until Bose stimulation causes it to take off at the onset time $t = \tau_{\mbox{\scriptsize{on}}}$. Since $f_0 (t)$ is very small for $t < \tau_{\mbox{\scriptsize{on}}}$, the three-body loss of condensate atoms is negligible at these early times. Thus the onset time is largely unaffected by the presence of three-body recombination. For $t > \tau_{\mbox{\scriptsize{on}}}$, $f_0 (t)$ grows rapidly, greatly increasing the particle density and enhancing the three-body loss rate. The condensate fraction attains its maximum value $F_0^{\mbox{\scriptsize{peak}}}$ at $t = \tau_{\mbox{\scriptsize{peak}}}$ when the condensate decay rate balances the rate of particle scattering into the condensate. Thereafter, $f_0 (t)$ decreases due to three-body loss. Since the condensate holds a macroscopic number of particles, the atom density in the condensate far exceeds that of any of the excited states. Therefore, $f_0 (t)$ decays much faster than either the non-condensate fraction in the dimple $f_{nc} (t)$ or the reservoir fraction $f_r (t)$. The preferential ejection of low-energy atoms via three-body recombination leads to evaporative heating. This heating, along with the particle loss, ultimately results in the death of the condensate. Thus we get a finite condensate lifetime, $\Delta t_{\mbox{\scriptsize{lf}}}$, defined as the duration for which $f_0 (t)$ is larger than half its maximum value. These general features are illustrated in Fig. \ref{f0loss} where we plot $f_0 (t)$ for a specific set of parameter values. In the following we discuss how $F_0^{\mbox{\scriptsize{peak}}}$, $\Delta t_{\mbox{\scriptsize{lf}}}$, $\tau_{\mbox{\scriptsize{peak}}}$, and $\tau_{\mbox{\scriptsize{on}}}$ vary with the different parameters.

\begin{figure}[h]
\includegraphics[width=\columnwidth]{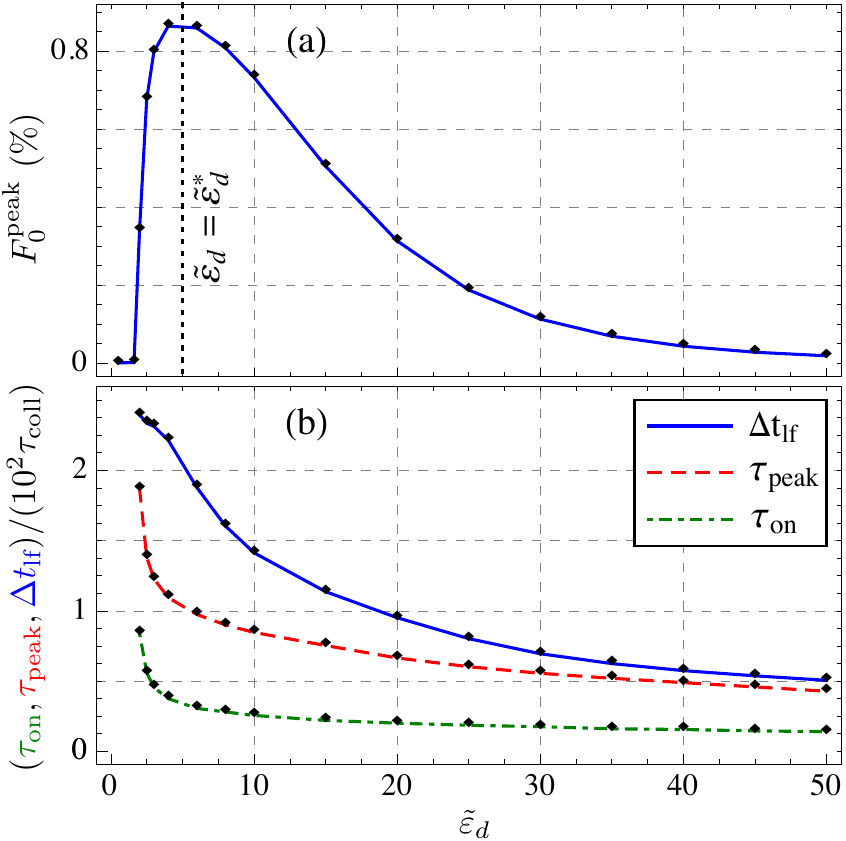}
\caption{(Color online) Variation of (a) the peak condensate fraction $F_0^{\mbox{\scriptsize{peak}}}$, (b) the onset time $\tau_{\mbox{\scriptsize{on}}}$ (dot-dashed green), the peak time $\tau_{\mbox{\scriptsize{peak}}}$ (dashed red), and the condensate lifetime $\Delta t_{\mbox{\scriptsize{lf}}}$ (solid blue) with the dimple depth $\tilde{\varepsilon}_d \equiv \beta_{r0} \varepsilon_d$ for $\Omega = 2000$, $\rho_i = 0.2$, and $T_{r0} = 0.1$ $\mu$K. For $\tilde{\varepsilon}_d < |\tilde{\mu}_{r0}|$, the atoms do not condense. As $\tilde{\varepsilon}_d$ is increased, the two-body scattering rate into the condensate grows rapidly at first, then falls off at large $\tilde{\varepsilon}_d$. The peak condensate fraction is reached when the three body loss rate balances the two-body scattering rate. Since the three-body rate depends only on the atom density, the non-monotonic variation of the two-body rate shows up in the variation of $F_0^{\mbox{\scriptsize{peak}}}$ with $\tilde{\varepsilon}_d$. The growth of the two-body rate above the condensation threshold decreases $\tau_{\mbox{\scriptsize{on}}}$ and $\tau_{\mbox{\scriptsize{peak}}}$. These vary little at large $\tilde{\varepsilon}_d$ since both $F_0^{\mbox{\scriptsize{peak}}}$ and the two-body rate fall off. At large $\tilde{\varepsilon}_d$, three-body loss is dominated by collisions between the non-condensed dimple atoms and the condensate atoms. The number of such non-condensed atoms grows with $\tilde{\varepsilon}_d$, yielding shorter lifetimes.}
\label{fpeakedimple}
\end{figure}

Figure \ref{fpeakedimple} shows the variation of the peak condensate fraction and the three timescales with the dimple depth $\tilde{\varepsilon}_d$ when other parameters are held fixed. Similar to Fig. \ref{F0}, we find that the variation of $F_0^{\mbox{\scriptsize{peak}}}$ with 
$\tilde{\varepsilon}_d$ is non-monotonic. However, the maximum value as well as the optimal dimple depth $\tilde{\varepsilon}_d^*$ are significantly reduced by three-body loss. When $\varepsilon_d < |\mu_{r0}|$, the initial chemical potential lies below the dimple bottom, and the atoms do not condense, so $F_0^{\mbox{\scriptsize{peak}}}$ vanishes. At larger dimple depths, the population of the condensate is governed by the competition between two-body collisions scattering particles into the condensate and three-body recombinations causing particles to leave the condensate. The condensate fraction reaches its peak when the two-body and three-body rates balance each other. As $\varepsilon_d$ is increased beyond the threshold $|\mu_{r0}|$, the two-body rate climbs as the phase space density increases, then falls due to increased heating and a reduced overlap between the initial and final states. The three-body rate only depends on the density, so the $F_0^{\mbox{\scriptsize{peak}}}$ vs $\tilde{\varepsilon}_d$ curve follows the variation of the two-body scattering rate.
The rapid increase of the two-body rate with $\varepsilon_d$ just above the threshold leads to a sharp decrease in the onset time, $\tau_{\mbox{\scriptsize{on}}}$, and the time required to reach peak condensate fraction, $\tau_{\mbox{\scriptsize{peak}}}$. At large dimple depths, $\tau_{\mbox{\scriptsize{peak}}}$ decrease very slowly because both $F_0^{\mbox{\scriptsize{peak}}}$ and the two-body rate fall off. We also find that the condensate lifetime $\Delta t_{\mbox{\scriptsize{lf}}}$ decreases monotonically with $\tilde{\varepsilon}_d$. For large $\tilde{\varepsilon}_d$, the condensate loss is dominated by collisions between the non-condensed dimple atoms and the condensate. The number of such non-condensed dimple atoms grows with $\tilde{\varepsilon}_d$, yielding the shorter lifetimes.

\begin{figure}[t]
\includegraphics[width=\columnwidth]{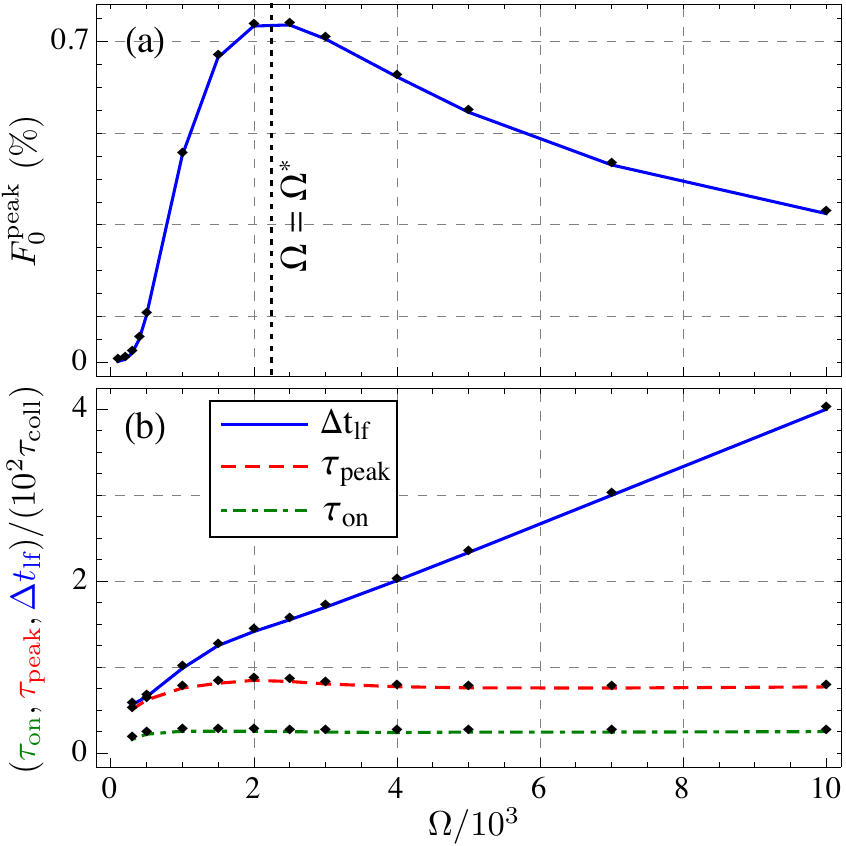}
\caption{(Color online) Variation of (a) $F_0^{\mbox{\scriptsize{peak}}}$, (b) $\tau_{\mbox{\scriptsize{on}}}$ (dot-dashed green), $\tau_{\mbox{\scriptsize{peak}}}$ (dashed red), and $\Delta t_{\mbox{\scriptsize{lf}}}$ (solid blue) with the volume ratio $\Omega$ for $\tilde{\varepsilon}_d = 10$, $\rho_i = 0.2$, and $T_{r0} = 0.1$ $\mu$K. When $\Omega$ is very small, the total atom number $\mathcal{N}$ is small and the dimple can hold its entire population in the excited states, so condensation does not occur. As $\Omega$ is increased, $F_0^{\mbox{\scriptsize{peak}}}$ grows rapidly at first, then falls off when $\Omega$ becomes larger than an optimal value $\Omega^*$. At large $\Omega$, the condensate population $N_0$ far exceeds the non-condensate population in the dimple, and three-body loss is dominated by collisions among the condensate atoms. Balancing the two-body growth rate and the three-body decay rate gives a peak condensate size $N_0^{\mbox{\scriptsize{peak}}}$ which is independent of $\Omega$. Since $\mathcal{N} \propto \Omega$, $F_0^{\mbox{\scriptsize{peak}}}$ falls off as $1/\Omega$. The condensate lifetime is set by the depletion rate of the reservoir. When $\Omega$ is large, the condensate size and hence the loss rate becomes independent of $\Omega$. Thus it takes longer to empty a larger reservoir, causing $\Delta t_{\mbox{\scriptsize{lf}}}$ to grow linearly. We also find that $\tau_{\mbox{\scriptsize{on}}}$ and $\tau_{\mbox{\scriptsize{peak}}}$ are mostly independent of $\Omega$.}
\label{fpeakomega}
\end{figure}

\begin{figure}[t]
\includegraphics[width=\columnwidth]{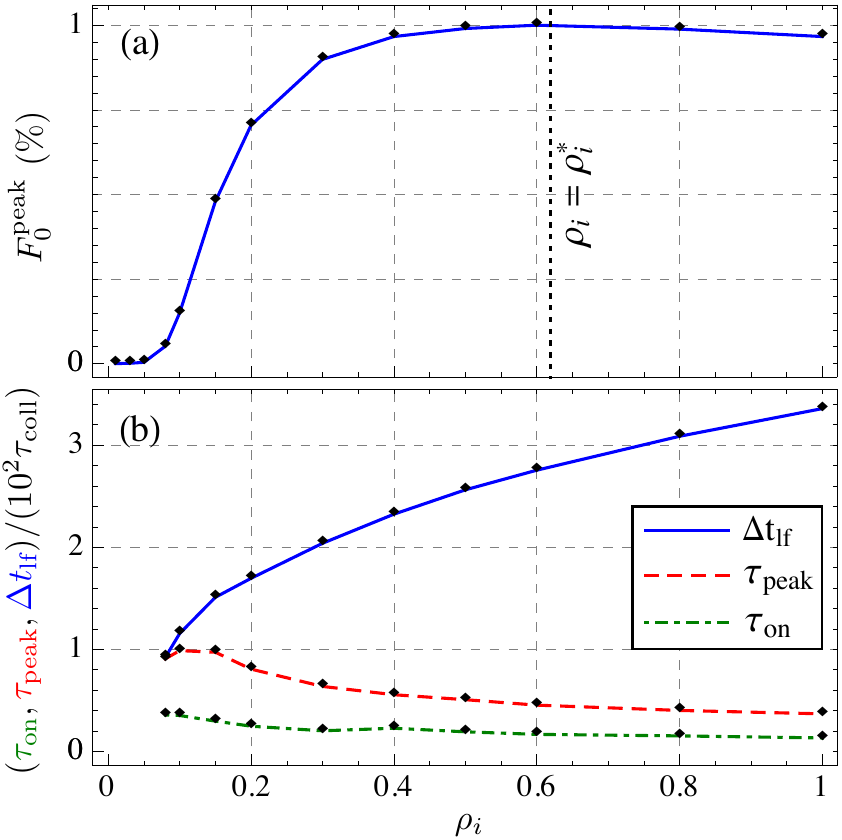}
\caption{(Color online) Variation of (a) $F_0^{\mbox{\scriptsize{peak}}}$, (b) $\tau_{\mbox{\scriptsize{on}}}$ (dot-dashed green), $\tau_{\mbox{\scriptsize{peak}}}$ (dashed red), and $\Delta t_{\mbox{\scriptsize{lf}}}$ (solid blue) with the initial phase space density $\rho_i$ for $\tilde{\varepsilon}_d = 10$, $\Omega = 3000$, and $T_{r0} = 0.1$ $\mu$K. When $\rho_i$ is very small, the atoms do not condense. As $\rho_i$ is increased, $F_0^{\mbox{\scriptsize{peak}}}$ rises rapidly at first, then falls at large $\rho_i$ because the three-body decay rate of the condensate fraction grows faster than the two-body growth rate. Thus we get an optimal phase space density $\rho_i^*$ which yields the largest condensate, although this peak is much less pronounced compared to the peaks seen when one varies $\tilde{\varepsilon}_d$ (Fig. \ref{fpeakedimple}) or $\Omega$ (Fig. \ref{fpeakomega}). The increased three-body decay rate at larger $\rho_i$ also leads to a smaller condensate lifetime: $\Delta t_{\mbox{\scriptsize{lf}}}$ falls off as $\rho_i^{-1/2}$ for large $\rho_i$. However, since $\tau_{\mbox{\scriptsize{coll}}}$ diminishes more rapidly as $1/\rho_i$, $\Delta t_{\mbox{\scriptsize{lf}}} / \tau_{\mbox{\scriptsize{coll}}}$ grows as $\sqrt{\rho_i}$. We find that $\tau_{\mbox{\scriptsize{on}}} / \tau_{\mbox{\scriptsize{coll}}}$ and $\tau_{\mbox{\scriptsize{peak}}} / \tau_{\mbox{\scriptsize{coll}}}$ decrease slowly with $\rho_i$.}
\label{fpeakrho}
\end{figure}

In Fig. \ref{fpeakomega}(a) we plot the variation of $F_0^{\mbox{\scriptsize{peak}}}$ with the volume ratio $\Omega$. Decreasing $\Omega$ reduces the total number of atoms ($\mathcal{N}$), without changing the maximum number of non-condensed atoms in the dimple. Therefore, when $\Omega$ becomes very small, the atoms in the dimple no longer condense, as seen experimentally in Ref. \cite{expdetails}. As $\Omega$ is increased, the condensate fraction increases rapidly until $\Omega$ reaches an optimal value $\Omega^*$, beyond which $F_0^{\mbox{\scriptsize{peak}}}$ falls off due to increased three-body loss. We see a similar variation of $F_0^{\mbox{\scriptsize{peak}}}$ with $\rho_i$ in Fig. \ref{fpeakrho}(a). When $\rho_i$ is very small, few atoms populate the dimple and no condensation takes place. As one increases $\rho_i$, $F_0^{\mbox{\scriptsize{peak}}}$ rises rapidly at first, then falls for $\rho_i > \rho_i^*$. The increase of the condensate fraction with $\Omega$ and $\rho_i$ was also seen in Fig. \ref{F0} where no inelastic loss was assumed. The fall-off at large $\Omega$ or large $\rho_i$ can be explained as follows: As $\Omega$ or $\rho_i$ is increased, the total particle number $\mathcal{N}$ and the condensate population $N_0$ grows, however, the number of non-condensed atoms in the dimple ($N_{nc}$) does not change. Therefore at large $\Omega$ or $\rho_i$, $N_0 \gg N_{nc}$ . The non-condensate population merely acts as a medium to transfer particles from the reservoir to the condensate. In addition, three-body loss principally occurs in the dimple, resulting from collisions among the densely packed condensate atoms. Thus the dynamics are governed by the simplified rate equations
\begin{equation}
\dot{N}_0 = c_1 n N_0 - c_2 N_0^3 \hspace{0.05cm}, \quad \mbox{and} \quad \dot{n} = -(c_1 / V_r) \hspace{0.03cm} n N_0 \hspace{0.05cm},
\label{omegarhoscaling}
\end{equation}
where $V_r$ is the reservoir volume and $n$ is the density of reservoir particles. 
The rate coefficients $c_1$ and $c_2$ represent two-particle collisions which fill the dimple, and three-body losses respectively. They depend on the reservoir temperature but do not depend explicitly on $\Omega$ or $\rho_i$. From our full model we find that the temperature does not vary much ($\sim 25 \%$) during the condensate lifetime, and depends very weakly on $\Omega$ and $\rho_i$ (also seen in Fig. \ref{Tf}). Thus we treat $c_1$ and $c_2$ as constants in solving Eq. (\ref{omegarhoscaling}). $N_0$ attains its peak value when $\dot{N}_0 = 0$. Thus $N_0^{\mbox{\scriptsize{peak}}} = (c_1 n(\tau_{\mbox{\scriptsize{peak}}}) / c_2)^{1/2}$. From Figs. \ref{fpeakomega}(b) and \ref{fpeakrho}(b) we see that when $\Omega$ or $\rho_i$ is large, the condensate reaches its peak size very quickly, then decays gradually. For $0 < t < \tau_{\mbox{\scriptsize{peak}}}$, the change in $n$ is negligible. Hence $N_0^{\mbox{\scriptsize{peak}}} \approx (c_1 n(0) / c_2)^{1/2} \propto \rho_i^{1/2}$, and $F_0^{\mbox{\scriptsize{peak}}} \equiv N_0^{\mbox{\scriptsize{peak}}} / \mathcal{N} = N_0^{\mbox{\scriptsize{peak}}} / (\rho_i \Omega \hspace{0.05cm}\tilde{l}_d^3) \propto \Omega^{-1} \rho_i^{-1/2}$. This accounts for the reduction of $F_0^{\mbox{\scriptsize{peak}}}$ at large $\Omega$ or $\rho_i$. To calculate the lifetime we note that the transfer of one atom from the reservoir to the condensate decreases $n$ by $1/V_r \ll n(0)$. Therefore, as the growth rate of $N_0$ declines after $t = \tau_{\mbox{\scriptsize{peak}}}$, the decay rate also falls to maintain $\dot{N}_0 \approx 0$, or $N_0 (t) \approx (c_1 n(t) / c_2)^{1/2}$. Thus the lifetime is set by the time required for the reservoir to be depleted \cite{expdetails}. Using the above expression for $N_0 (t)$ in the other equation, we find $1 / N_0 (t) = 1 / N_0^{\mbox{\scriptsize{peak}}} + c_1 t / 2 V_r$, which gives a lifetime $\Delta t_{\mbox{\scriptsize{lf}}} \approx 2 V_r/ (c_1 N_0^{\mbox{\scriptsize{peak}}}) \propto \Omega \hspace{0.02cm}\rho_i^{-1/2}$. Since $\tau_{\mbox{\scriptsize{coll}}}$ falls off with $\rho_i$ as $1/\rho_i$ (Eq. (\ref{taucoll})), $\Delta t_{\mbox{\scriptsize{lf}}} / \tau_{\mbox{\scriptsize{coll}}} \propto \Omega \hspace{0.02cm}\rho_i^{1/2}$. Such variation of the lifetime is illustrated in Figs. \ref{fpeakomega}(b) and \ref{fpeakrho}(b). We find that $\tau_{\mbox{\scriptsize{on}}}$ and $\tau_{\mbox{\scriptsize{peak}}}$ vary little with $\Omega$. As was true without three-body collisions, $\tau_{\mbox{\scriptsize{on}}} / \tau_{\mbox{\scriptsize{coll}}}$ decreases slowly with $\rho_i$. $\tau_{\mbox{\scriptsize{peak}}} / \tau_{\mbox{\scriptsize{coll}}}$ also falls with $\rho_i$, following the variation of $\tau_{\mbox{\scriptsize{on}}} / \tau_{\mbox{\scriptsize{coll}}}$.

\begin{figure}[t]
\includegraphics[width=\columnwidth]{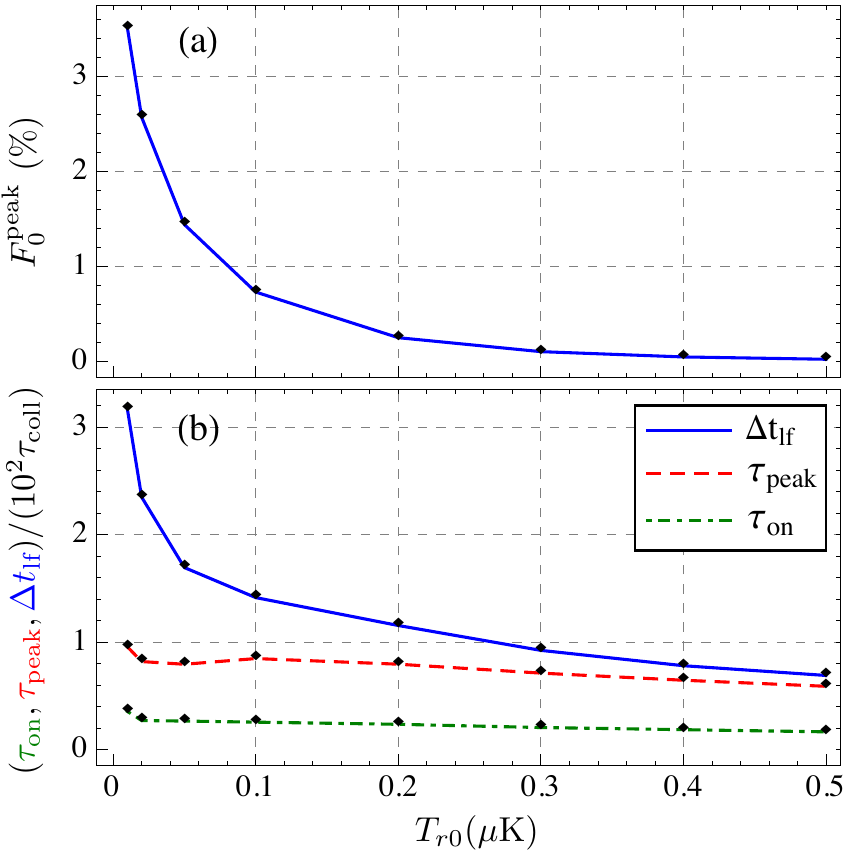}
\caption{(Color online) Variation of (a) $F_0^{\mbox{\scriptsize{peak}}}$, (b) $\tau_{\mbox{\scriptsize{on}}}$ (dot-dashed green), $\tau_{\mbox{\scriptsize{peak}}}$ (dashed red), and $\Delta t_{\mbox{\scriptsize{lf}}}$ (solid blue) with the initial reservoir temperature $T_{r0}$ for $\tilde{\varepsilon}_d = 10$, $\Omega = 2000$, and $\rho_i = 0.2$. The rate of two-body collisions which populate the dimple is set by $1/\tau_{\mbox{\scriptsize{coll}}} \propto T_{r0}^2$, whereas the three-body loss rate grows as $T_{r0}^3$. Thus a higher $T_{r0}$ increases the three-body loss rate relative to the two-body scattering rate, causing both $F_0^{\mbox{\scriptsize{peak}}}$ and $\Delta t_{\mbox{\scriptsize{lf}}} / \tau_{\mbox{\scriptsize{coll}}}$ to decrease. $\Delta t_{\mbox{\scriptsize{lf}}} / \tau_{\mbox{\scriptsize{coll}}}$ diverges as $1/T{r0}$ for small $T_{r0}$. Since $F_0^{\mbox{\scriptsize{peak}}}$ is smaller at larger $T_{r0}$, it takes fewer two-body collisions to reach this value, so $\tau_{\mbox{\scriptsize{peak}}} / \tau_{\mbox{\scriptsize{coll}}}$ decreases slowly with $T_{r0}$. The onset of condensation is not affected much by three-body loss, thus $\tau_{\mbox{\scriptsize{on}}} / \tau_{\mbox{\scriptsize{coll}}}$ is nearly independent of $T_{r0}$.}
\label{fpeaktemp}
\end{figure}

Figure \ref{fpeaktemp} shows how $F_0^{\mbox{\scriptsize{peak}}}$, $\Delta t_{\mbox{\scriptsize{lf}}}$, $\tau_{\mbox{\scriptsize{peak}}}$, and $\tau_{\mbox{\scriptsize{on}}}$ vary with the initial reservoir temperature $T_{r0}$. To understand the features, we note that the growth and redistribution of the dimple population occur via two-body collisions. Thus the rates of these processes are set by $1/\tau_{\mbox{\scriptsize{coll}}}$ which is proportional to $T_{r0}^2$ (see Eqs. (\ref{rhoi}) and (\ref{taucoll})). However, from Eq. (\ref{f0l}) we find that the three-body loss rate is proportional to $T_{r0}^3$. Therefore, a higher initial temperature increases the strength of three-body decay processes relative to two-body elastic processes. This reduces the peak condensate fraction as well as the condensate lifetime. We find that $\Delta t_{\mbox{\scriptsize{lf}}} / \tau_{\mbox{\scriptsize{coll}}}$ diverges as $1/T_{r0}$ for small $T_{r0}$. Since $F_0^{\mbox{\scriptsize{peak}}}$ is smaller for larger $T_{r0}$, it takes fewer two-body collisions to reach the peak condensate fraction. Consequently, $\tau_{\mbox{\scriptsize{peak}}} / \tau_{\mbox{\scriptsize{coll}}}$ decreases slowly with $T_{r0}$. We also notice that $\tau_{\mbox{\scriptsize{on}}} / \tau_{\mbox{\scriptsize{coll}}}$ stays essentially constant as $T_{r0}$ is varied because the onset of condensation is governed by two-body processes alone.

\subsection{Effect of finite trap depth}\label{finite trap depth}

Here we discuss how the above results are altered when the reservoir trap has a finite depth $\varepsilon_t$. First we remind the reader that we have modeled the growth and redistribution of the dimple population by two kinds of elastic collisions, as illustrated in Fig. \ref{processes}. Both processes can be either one-way or two-way: if the recoiling reservoir atom has a total energy greater than $\varepsilon_t$, it escapes from the trap. Such a collision has no reverse process and happens only one-way. Other collisions happen both ways.

\begin{figure}[h]
\includegraphics[width=\columnwidth]{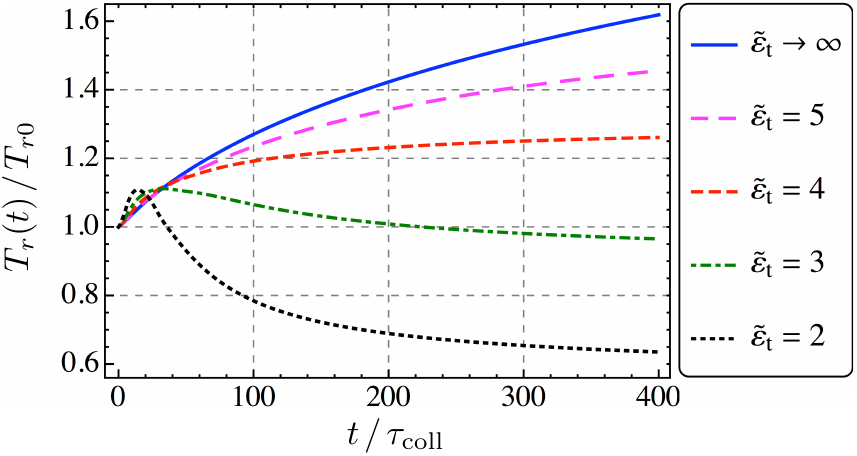}
\caption{(Color online) Time-evolution of the reservoir temperature $T_r$ for different values of the trap depth $\tilde{\varepsilon}_t \equiv \beta_{r0} \varepsilon_t$ with $\tilde{\varepsilon}_d = 10$, $\Omega = 2000$, $\rho_i = 0.2$, and $T_{r0} = 0.1$ $\mu$K. When $\tilde{\varepsilon}_t \to \infty$ (solid blue line), particle transfer from the reservoir to the dimple and their redistribution from the higher-energy to the lower-energy dimple states heat the reservoir. Along with evaporative heating by three-body loss, this causes $T_r$ to rise monotonically. When $\tilde{\varepsilon}_t$ is finite (dashed and dotted lines), a reservoir atom can recoil from a collision with a total energy greater than $\varepsilon_t$ and leave the trap. The average initial energy of this atom is less than the average particle energy in the reservoir since the dimple is located at the trap center. Therefore such one-way collisions contribute to heating at short times. When the atom density in the dimple becomes sufficiently large, both one-way and two-way collisions initiate thermalization by transferring atoms from the higher to the lower-energy dimple states. However, the one-way transfer soon results in an excess of particles in the low-energy states. Two-way collisions now transfer these extra particles to higher-energy states and thus cool the reservoir. The cooling rate increases as $\tilde{\varepsilon}_t$ is lowered. Thus lower trap depths yield lower final temperatures.}
\label{Trloss}
\end{figure}

When $\tilde{\varepsilon}_t \equiv \beta_{r0} \varepsilon_t \to \infty$, only two-way collisions are present. After the dimple is turned on, two-way growth processes start populating the dimple. Such processes reduce the number of reservoir atoms $N_r$, but increase their total energy $E_r$, thus heating up the reservoir. When the atom density in the dimple becomes comparable to that in the reservoir, two-way redistribution processes transfer atoms to the lower energy dimple states, leading to thermalization (see Fig. \ref{thermalization}). These redistribution processes do not change $N_r$, but increase $E_r$, causing heating. Three-body recombinations also cause evaporative heating. Thus the reservoir temperature $T_r$ increases monotonically, as shown by the solid blue curve in Fig. \ref{Trloss}. When $\tilde{\varepsilon}_t$ is finite, both one-way and two-way collisions are present. In a one-way growth process, the colliding atoms are removed from the reservoir. Since the dimple is located at the trap center, the average energy of a colliding atom is less than the average energy per particle in the reservoir. Therefore, one-way growth processes (and for the same reason, one-way redistribution processes) contribute to heating. Thus we find that $T_r$ always increases just after turning on the dimple. When the atom density in the dimple becomes large enough, redistribution processes start operating. At first, both one-way and two-way collisions cause a net transfer of atoms from the higher energy to the lower energy dimple states. However, the one way particle transfer soon overcompensates thermalization, resulting in an excess of atoms in the low-energy states. This imbalance flips the direction of the two-way traffic, which now transfers atoms to the higher energy states. Such two-way collisions decrease $E_r$ without changing $N_r$, thus cooling the reservoir. A smaller $\tilde{\varepsilon}_t$ results in a larger imbalance of the atom distribution, which increases the cooling rate. Thus we see in Fig. \ref{Trloss} that $T_r$ decreases after the initial growth when $\tilde{\varepsilon}_t$ is sufficiently small compared to $\tilde{\varepsilon}_d$.

\begin{figure}[h]
\includegraphics[width=\columnwidth]{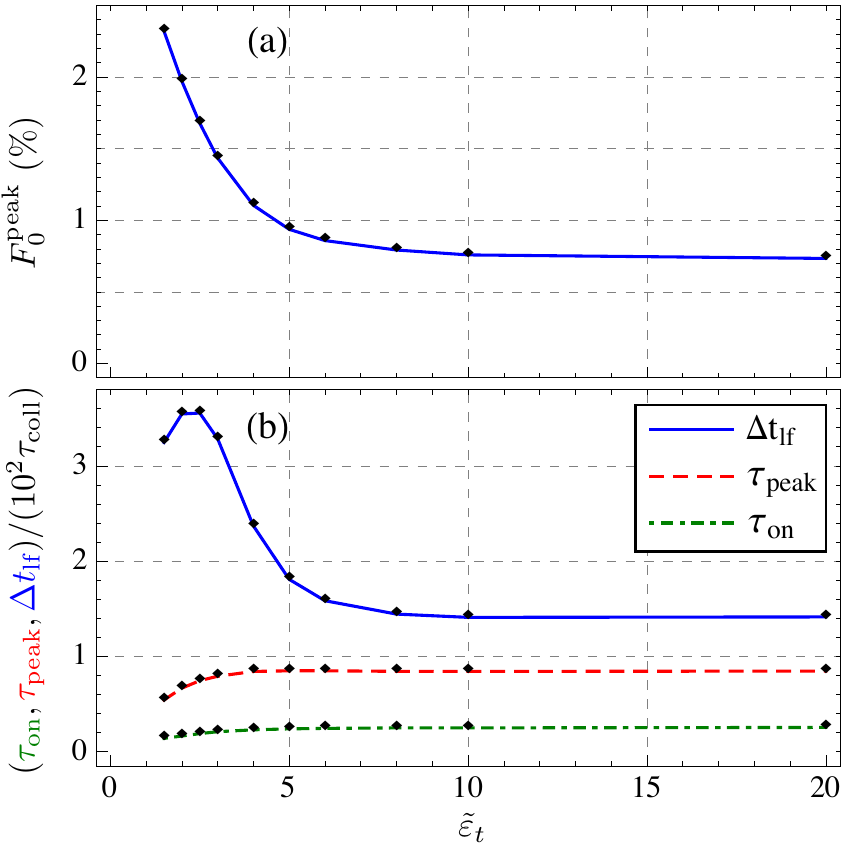}
\caption{(Color online) Variation of (a) $F_0^{\mbox{\scriptsize{peak}}}$, (b) $\tau_{\mbox{\scriptsize{on}}}$ (dot-dashed green), $\tau_{\mbox{\scriptsize{peak}}}$ (dashed red), and $\Delta t_{\mbox{\scriptsize{lf}}}$ (solid blue) with the trap depth $\tilde{\varepsilon}_t \equiv \beta_{r0} \varepsilon_t$ for $\tilde{\varepsilon}_d = 10$, $\Omega = 2000$, $\rho_i = 0.2$, and $T_{r0} = 0.1$ $\mu$K. As $\tilde{\varepsilon}_t$ is decreased, the rate of one-way redistribution processes which transfer atoms from the higher-energy to the lower-energy dimple states increases. This gives rise to a faster growth of the condensate, which increases $F_0^{\mbox{\scriptsize{peak}}}$ and decreases $\tau_{\mbox{\scriptsize{on}}}$ and $\tau_{\mbox{\scriptsize{peak}}}$. A smaller trap depth also causes more cooling, which reduces the non-condensate fraction $f_{nc}$ and thus the three-body decay rate of the condensate. Hence we get a larger condensate lifetime. However, when $\tilde{\varepsilon}_t$ becomes very small, the increased evaporation rate of the reservoir limits the growth of $\Delta t_{\mbox{\scriptsize{lf}}} / \tau_{\mbox{\scriptsize{coll}}}$.}
\label{fpeaketrap}
\end{figure}

In Fig. \ref{fpeaketrap} we plot the variation of the peak condensate fraction $F_0^{\mbox{\scriptsize{peak}}}$ and the timescales with the trap depth $\tilde{\varepsilon}_t$. As $\tilde{\varepsilon}_t$ is decreased, one-way redistribution processes become stronger, leading to a faster growth of the condensate. Consequently, a higher condensate density has to be reached before the three-body decay rate can balance the growth rate. Thus $F_0^{\mbox{\scriptsize{peak}}}$ grows as the trap is made shallower. The larger growth rate also reduces the onset time $\tau_{\mbox{\scriptsize{on}}}$ and the time required to reach the peak, $\tau_{\mbox{\scriptsize{peak}}}$. A smaller trap depth causes less heating and reduces the non-condensate fraction $f_{nc}$, which in turn decreases the three-body decay rate of the condensate. Combined with the faster growth rate, this yields a larger condensate lifetime $\Delta t_{\mbox{\scriptsize{lf}}}$. However, as the trap depth is lowered, the evaporation rate of the reservoir atoms also increases, and eventually becomes comparable to the decay rate of the condensate. As the number of reservoir atoms falls, so does the rate of particle transfer from the reservoir to the dimple, and hence the growth rate of the condensate. This limits the rise of $\Delta t_{\mbox{\scriptsize{lf}}} / \tau_{\mbox{\scriptsize{coll}}}$. Additionally, $\tau_{\mbox{\scriptsize{coll}}}$ grows at small $\tilde{\varepsilon}_t$ (Eq. (\ref{taucoll})), causing $\Delta t_{\mbox{\scriptsize{lf}}} / \tau_{\mbox{\scriptsize{coll}}}$ to decrease.

\section{Summary and Outlook}\label{summary}

In this work we have studied the condensation kinetics of weakly interacting Bosons in a dimple potential using quantum kinetic rate equations. We have modeled the growth and redistribution of the dimple population by two-body elastic collisions. We have incorporated three-body inelastic losses for ${}^{87}$Rb, and varied the reservoir trap depth $\varepsilon_t$ to study the effects of evaporation. The dynamics are controlled by the dimple depth $\varepsilon_d$, the ratio of the reservoir volume to the dimple volume $\Omega$, the initial phase space density $\rho_i$, and the initial temperature $T_{r0}$. The absolute size of the dimple does not matter as long as it is much larger than the thermal wavelength. We have presented detailed results for condensate fraction, the temperature, and the different timescales. Our results are consistent with features observed in recent experiments, and should provide a useful guide for more efficient production of condensates in the future.

We find that the initial growth of the dimple population is dominated by states whose energy is near half the dimple depth. However, scattering between levels quickly transfers these particles to the low-energy states, giving rise to a bimodal particle distribution at $t \approx 2 \tau_{\mbox{\scriptsize{coll}}}$ when $\varepsilon_d \gtrsim 8 k_B T_{r0}$. The dimple attains quasi-thermal equilibrium in about $8 \tau_{\mbox{\scriptsize{coll}}}$ after it is turned on. Comparable thermalization timescales were reported in previous studies \cite{ketterle98evaporative, esslinger07, walraven96, gardiner97trap, expdetails, wolfe89, kagan92}. The condensate grows slowly at first until Bose stimulation can take over. This results in a time-delay $\tau_{\mbox{\scriptsize{on}}}$ before the onset of condensation, as was seen in Refs. \cite{davis11, grimm13}. When $\varepsilon_t \gg \varepsilon_d$ and $k_B T_{r0}$, particle scattering into the dimple causes heating in the reservoir. In the absence of three-body loss, the dimple population saturates at a value limited by this heating. The saturation time is proportional to both $\tau_{\mbox{\scriptsize{coll}}}$ and $\Omega$, and increases monotonically with $\varepsilon_d$. When $\varepsilon_d$ is small such that the initial chemical potential is below the dimple bottom, we do not get condensation. As $\varepsilon_d$ is increased beyond this threshold, the saturation condensate fraction $F_0$ grows rapidly at first, then falls off due to increased heating. This gives rise to an optimal dimple depth $\varepsilon_d^*$ which yields the largest condensate fraction. The onset time $\tau_{\mbox{\scriptsize{on}}}$ is also minimized at $\varepsilon_d = \varepsilon_d^*$. We find that $\varepsilon_d^*$ scales with $k_B T_{r0}$. The non-monotonic behavior of the condensate fraction was observed in a recent experiment \cite{davis11}. A larger $\rho_i$ or $\Omega$ both favor condensation, increasing $F_0$ and $\varepsilon_d^*$, while decreasing $\tau_{\mbox{\scriptsize{on}}}$. The reduction of $\tau_{\mbox{\scriptsize{on}}}$ was seen in Ref. \cite{expdetails}. The dynamics become more non-adiabatic at larger $\varepsilon_d$, with the entropy growing by $20 \%$ when $\varepsilon_d \approx 20 k_B T_{r0}$. Typical experiments have $\varepsilon_d$ ranging from a few $k_B T_{r0}$ to $\sim \hspace{-0.05cm} 10 k_B T_{r0}$ \cite{davis11, grimm13}.

We find that three-body loss plays an important role for ${}^{87}$Rb, reducing the maximum condensate fraction to a few percent for $T_{r0} = 100$ nK. It also limits the condensate lifetime $\Delta t_{\mbox{\scriptsize{lf}}}$. The condensate fraction now decays toward zero after reaching a peak value $F_0^{\mbox{\scriptsize{peak}}}$ at $t = \tau_{\mbox{\scriptsize{peak}}}$. $F_0^{\mbox{\scriptsize{peak}}}$ varies non-monotonically with $\varepsilon_d$ similar to $F_0$. However, both its maximum value and the optimal dimple depth $\varepsilon_d^*$ are significantly reduced by three-body loss, in agreement with similar modeling in Ref. \cite{davis11}. When $\rho_i$ or $\Omega$ is very small, condensation does not occur, as seen experimentally in Ref. \cite{grimm13}. As $\rho_i$ or $\Omega$ is increased, $F_0^{\mbox{\scriptsize{peak}}}$ grows rapidly at first, then falls due to increased loss rate resulting from higher local densities. Thus there exist an optimal volume ratio $\Omega^*$ and an optimal initial phase space density $\rho_i^*$ which yield maximum condensate fraction. We find that the peak at $\rho_i = \rho_i^*$ is much flatter than either of those at $\Omega = \Omega^*$ or $\varepsilon_d = \varepsilon_d^*$. The three-body decay rate grows much faster with $T_{r0}$ than the rate of two-body collisions, causing $F_0^{\mbox{\scriptsize{peak}}}$ to drop. We find that $\Delta t_{\mbox{\scriptsize{lf}}} / \tau_{\mbox{\scriptsize{coll}}}$ increases with $\rho_i$ and $\Omega$, and decreases with $\varepsilon_d$ and $T_{r0}$. We also find that $\tau_{\mbox{\scriptsize{peak}}}$ follows the variation of $\tau_{\mbox{\scriptsize{on}}}$, falling off with $\varepsilon_d$, $\rho_i$, and $T_{r0}$, while being almost independent of $\Omega$. When the trap depth is finite, particles recoiling with sufficiently high energies from elastic collisions escape from the trap. This leads to cooling. Lower trap depths yield lower final temperatures and enhance the condensate growth rate, producing larger and longer-lived condensates. However, at very small trap depths, the increased evaporation rate of the reservoir limits the condensate lifetime.

Several of our predictions are amenable to testing in future experiments. The bimodal shape of $f(\tilde{E},t)$ at $t \approx 2 \tau_{\mbox{\scriptsize{coll}}}$ should show up in time-of-flight images as two expanding shells of atoms, though their actual shape would depend on the trap geometry. Our predictions for condensate fractions can readily be checked using the techniques in Refs. \cite{ketterle98dimple, grimm03, dalibard11, davis11, grimm13}.

Our model can be readily generalized to study experiments where the dimple is turned on gradually, making the loading process more adiabatic \cite{pinkse97, ketterle98dimple, grimm03, davis11}. A gradual turn-on is likely to increase the condensate fraction, although recent experiments suggest that it does not affect the dynamics at large times \cite{expdetails}. One can also make the reservoir trap anisotropic \cite{davis11, grimm13}, or vary the location of the dimple \cite{uncu0708}. Dimples located off-center in a harmonic trap should have smaller filling rates due to lower particle density, but can assist in evaporative cooling since the particles would have higher energies. In Ref. \cite{grimm13}, Stellmer {\it et al.} employed a novel technique where the reservoir is continuously laser cooled while the dimple particles are rendered transparent to the cooling photons by a blue-detuned laser beam. This prevents heating of the reservoir and significantly increases the condensate lifetime. It would be valuable to study how this technique alters the kinetics in future theory work. One could also explore the loading of arrays of dimples. Our framework can be naturally extended to model such experiments. Dimple methods can also be applied to Fermions \cite{bernier09} or Boson-Fermion mixtures.

In modeling the kinetics, we have made a few simplifying assumptions to reduce the computational complexity. In particular, we have not included mean-field interactions between the condensate and the thermal cloud \cite{davis11, stoof01, stoof00, gardiner98, gardiner00constantbath, gardiner00changingbath, holland99, aspect04, cornell97, dalfovo99, expdetails}, and we have neglected two kinds of elastic collisions, as described in the last paragraph in Sec. \ref{intro}. These can be incorporated in future refinements of our model. They might alter some quantitative predictions by factors of 2, but we do not expect them to change any of the qualitative features \cite{pinkse97, gardiner98, holland97, gardiner00constantbath}.

\section*{Acknowledgments}\label{acknowledgments}
We thank Mukund Vengalattore and his students for illuminating discussions. This work was supported by the National Science Foundation under Grant No. PHY-1068165, and the ARO-MURI Non-equilibrium Many-body Dynamics grant (63834-PH-MUR).

\appendix
\section{Rate equations for the growth of the dimple population}\label{growthapp}

The rate of inflow of particles to the $\vec{n}$-th dimple state is given by (Eq. (\ref{gin}))
\begin{align}
\hspace{-0.2cm} \nonumber \bigg(\frac{dN_{\vec{n}}}{dt}\bigg)^g_{in}& = \frac{2 \pi}{\hbar} U_0^2 z_r^2 (1 + N_{\vec{n}}) \int^{\prime} \hspace{-0.05cm} \frac{d^3 p \hspace{0.05cm} d^3 q}{(2 \pi \hbar)^6} \; e^{-\beta_r (\frac{p^2+q^2}{2m})}\\
\times \hspace{0.05cm} \delta & \bigg(\frac{p^2 + q^2 - \big(\vec{p} + \vec{q} - (2 \pi \hbar/l_d) \vec{n}\big)^2}{2m} - \varepsilon_{n}\bigg), \label{ginapp}
\end{align}
where the prime on the integral symbol denotes the condition that the initial momenta of the colliding particles must satisfy $p^2, q^2 < 2 m \varepsilon_t$. To simplify Eq. (\ref{ginapp}) we write it in terms of $\vec{\tilde{p}},\vec{\tilde{q}} \equiv (\beta_{r}/4m)^{1/2} (\vec{p} \pm \vec{q})$ and use the dispersion of the dimple modes, Eq. (\ref{energyntilde}). This gives
\begin{align}
\hspace{-0.2cm} \nonumber \bigg(\frac{dN_{\vec{n}}}{dt}\bigg)^g_{in} = \frac{2 m a^2 z_r^2}{\pi^3 \hbar^3 \beta_r^2} (1 + N_{\vec{n}}) \int^{\prime} \hspace{-0.1cm} d^3 \tilde{p} \hspace{0.05cm} d^3 \tilde{q} \; e^{-(\tilde{p}^2 + \tilde{q}^2)} &\\
\times \hspace{0.05cm} \delta\big( \tilde{q}^2 + \beta_r \varepsilon_d - \tilde{p}^2 - 2 \beta_r E_n + 2 (2 \beta_r E_n)^{1/2} \hspace{0.05cm} \vec{\tilde{p}}.\hat{n} \big)&, \label{ginsub}
\end{align}
where $\hat{n} \equiv \vec{n}/n$ and in the new variables, the prime stands for the constraint $\tilde{p}^2 + \tilde{q}^2 \pm 2 \vec{\tilde{p}}.\vec{\tilde{q}} < 2 \beta_r \varepsilon_t$. Dividing the integration region into separate parts we can express Eq. (\ref{ginsub}) as
\begin{equation}
\hspace{-0.25cm} \bigg(\frac{dN_{\vec{n}}}{dt}\bigg)^g_{in} \hspace{-0.05cm} = \tilde{G} \frac{\tilde{z}_r^2}{\tilde{\beta}_r^2} (1 + N_{\vec{n}}) \hspace{-0.05cm} \sum_{i=1,2} \hspace{-0.1cm} \mathscr{G}_i (\tilde{\beta}_r \tilde{\varepsilon}_t, \tilde{\beta}_r \tilde{\varepsilon}_d, \tilde{\beta}_r \tilde{E}_{n})\hspace{0.05cm}, \label{ginsimplified}
\end{equation}
where $\tilde{G} \equiv 16 \sqrt{2} \;(a/\lambda_{r0})^2 (z_{r0}^2 / \beta_{r0} \hbar)$ and
\begin{align}
\nonumber \mathscr{G}_1 (\tilde{\varepsilon}_t, \tilde{\varepsilon}_d, \tilde{E}) &\equiv \frac{1}{\sqrt{\hspace{-0.06cm} \tilde{E}}} \Bigg[\hspace{0.1cm} \iint\limits_{IA,C} d \tilde{p}\hspace{0.05cm} d \tilde{q}\hspace{0.05cm} \tilde{p}\hspace{0.05cm} \tilde{q}^2 \hspace{0.05cm} e^{-(\tilde{p}^2 + \tilde{q}^2)}\\
&\hspace{-0.6cm} + \iint\limits_{IB,C} d \tilde{p}\hspace{0.05cm} d \tilde{q}\hspace{0.05cm} \tilde{q} \hspace{0.05cm} \bigg(\tilde{\varepsilon}_t - \frac{\tilde{p}^2 + \tilde{q}^2}{2}\bigg) \hspace{0.05cm} e^{-(\tilde{p}^2 + \tilde{q}^2)} \Bigg], \label{g1}\\
\mathscr{G}_2 (\tilde{\varepsilon}_t, \tilde{\varepsilon}_d, \tilde{E}) &\equiv \frac{1}{\sqrt{\hspace{-0.06cm} \tilde{E}}} \hspace{0.05cm} \iint\limits_{II,C} d \tilde{p}\hspace{0.05cm} d \tilde{q}\hspace{0.05cm} \tilde{p}\hspace{0.05cm} \tilde{q}^2 \hspace{0.05cm} e^{-(\tilde{p}^2 + \tilde{q}^2)}\;. \label{g2}
\end{align}
Here the labels under the integrals denote the following conditions on $\tilde{p}$ and $\tilde{q}$:
\begin{eqnarray*}
IA &:& \tilde{p}^2 + \tilde{q}^2 \geq \tilde{\varepsilon}_t - \tilde{\varepsilon}_d + \tilde{E} \;\mbox{and}\; \tilde{p} + \tilde{q} < (2 \tilde{\varepsilon}_t)^{1/2} \label{IA}\\
IB &:& \tilde{p} + \tilde{q} \geq (2 \tilde{\varepsilon}_t)^{1/2} \;\mbox{and}\; \tilde{p}^2 + \tilde{q}^2 < 2 \tilde{\varepsilon}_t \label{IB}\\
II &:& \tilde{p}^2 + \tilde{q}^2 < \tilde{\varepsilon}_t - \tilde{\varepsilon}_d + \tilde{E} \label{II}\\
C &:& (\tilde{p} - (2 \tilde{E})^{1/2})^2 \leq \tilde{q}^2 + \tilde{\varepsilon}_d \leq (\tilde{p} + (2 \tilde{E})^{1/2})^2 \hspace{0.75cm} \label{C}
\end{eqnarray*}
Conditions $IA$ and $IB$ correspond to the range of initial momenta for which the atom recoiling back to the reservoir gains sufficient energy from the collision to escape from the trap. Such collisions happen only one-way: they do not have any reverse process. Whereas if condition $II$ is satisfied, no atom is lost from the trap, giving rise to two-way collisions. Condition $C$ ensures that both momentum and energy are conserved in the process.

Similarly, we simplify Eq. (\ref{gout}) describing the rate of particle flow out of the $\vec{n}$-th dimple state to obtain
\begin{equation}
\hspace{-0.05cm}\bigg(\frac{dN_{\vec{n}}}{dt}\bigg)^g_{out} \hspace{-0.35cm}= \hspace{-0.05cm} - \tilde{G} \frac{\tilde{z}_r}{z_{r0} \tilde{\beta}_r^2} N_{\vec{n}} \hspace{0.05cm} e^{-\tilde{\beta}_r (\tilde{\varepsilon}_d - \tilde{E}_{n})} \mathscr{G}_2 (\tilde{\beta}_r \tilde{\varepsilon}_t, \tilde{\beta}_r \tilde{\varepsilon}_d, \tilde{\beta}_r \tilde{E}_{n}) \hspace{0.025cm}. \label{goutsimplified}
\end{equation}
The net growth rate of $N_{\vec{n}} (t)$ is then found by adding Eqs. (\ref{ginsimplified}) and (\ref{goutsimplified}). We write this as a sum of contributions from one-way and two-way collisions:
\begin{equation}
\bigg(\frac{dN_{\vec{n}}}{dt}\bigg)^g = \bigg(\frac{dN_{\vec{n}}}{dt}\bigg)^g_{1} + \bigg(\frac{dN_{\vec{n}}}{dt}\bigg)^g_{2}\;,
\label{ngapp}
\end{equation}
where
\begin{align}
\bigg(\frac{dN_{\vec{n}}}{dt}\bigg)^g_{1} =&\hspace{0.1cm} \tilde{G} \frac{\tilde{z}_r^2}{\tilde{\beta}_r^2} (1 + N_{\vec{n}}) \hspace{0.05cm} \mathscr{G}_1 (\tilde{\beta}_r \tilde{\varepsilon}_t, \tilde{\beta}_r \tilde{\varepsilon}_d, \tilde{\beta}_r \tilde{E}_{n}) \hspace{0.05cm}, \label{ng1}\\
\nonumber \bigg(\frac{dN_{\vec{n}}}{dt}\bigg)^g_{2} =&\hspace{0.1cm} \tilde{G} \frac{\tilde{z}_r^2}{\tilde{\beta}_r^2} \Big[1 - N_{\vec{n}} \Big\{\frac{1}{z_{r}} e^{\tilde{\beta}_r (\tilde{\varepsilon}_d - \tilde{E}_{n})} - 1\Big\}\Big]\\
& \times \mathscr{G}_2 (\tilde{\beta}_r \tilde{\varepsilon}_t, \tilde{\beta}_r \tilde{\varepsilon}_d, \tilde{\beta}_r \tilde{E}_{n}) \hspace{0.05cm}. \label{ng2}
\end{align}

Each one-way collision reduces the number of atoms in the reservoir ($N_r$) by 2, whereas every two-way collision changes $N_r$ by 1. Hence we write
\begin{equation}
\bigg(\frac{dN_r}{dt}\bigg)^g = -\sum_{\vec{n}} \bigg[2 \hspace{0.05cm} \bigg(\frac{dN_{\vec{n}}}{dt}\bigg)^g_{1} + \bigg(\frac{dN_{\vec{n}}}{dt}\bigg)^g_{2} \hspace{0.05cm} \bigg].
\label{nrgapp}
\end{equation}

In a one-way collision, the total energy of the colliding particles is lost from the reservoir. Therefore, the rate at which one-way collisions decrease the total energy in the reservoir ($E_r$) can be written as (Eq. (\ref{Erg1n}))
\begin{align}
\nonumber \hspace{-0.05cm}\bigg(\frac{dE_{r}}{dt}\bigg)^g_{1} \hspace{-0.05cm} = -\frac{2 \pi}{\hbar} U_0^2 z_r^2 \sum_{\vec{n}} (1 + N_{\vec{n}}) \hspace{-0.05cm} \int^{\prime\prime} \hspace{-0.05cm} \frac{d^3 p \hspace{0.05cm} d^3 q}{(2 \pi \hbar)^6} \; e^{-\beta_r (\frac{p^2+q^2}{2m})}\\
\times \hspace{0.05cm}\frac{p^2 + q^2}{2m}\hspace{0.1cm}\delta\bigg(\frac{p^2 + q^2 - \big(\vec{p} + \vec{q} - (2 \pi \hbar/l_d) \vec{n}\big)^2}{2m} - \varepsilon_{n}\bigg), \label{Erg1app1}
\end{align}
where the double prime restricts the initial momenta to regions where $p^2,q^2 < 2 m \varepsilon_t$ and $p^2 + q^2 > 2 m  (\varepsilon_t - \varepsilon_d + E_n)$. To simplify Eq. (\ref{Erg1app1}) we apply the same operations as we did on Eq. (\ref{ginapp}). This yields
\begin{equation}
\bigg(\frac{dE_r}{dt}\bigg)^g_{1} = - \tilde{G} \frac{\tilde{z}_r^2}{\beta_r \tilde{\beta}_r^2} \sum_{\vec{n}} (1 + N_{\vec{n}}) \hspace{0.05cm} \mathscr{C}_g (\tilde{\beta}_r \tilde{\varepsilon}_t, \tilde{\beta}_r \tilde{\varepsilon}_d, \tilde{\beta}_r \tilde{E}_{n}) \hspace{0.05cm}, \label{Erg1app2}
\end{equation}
where
\begin{align}
\hspace{-0.15cm} \nonumber \mathscr{C}_g (\tilde{\varepsilon}_t, \tilde{\varepsilon}_d, \tilde{E}) &\equiv \frac{1}{\sqrt{\hspace{-0.06cm} \tilde{E}}} \Bigg[\hspace{0.1cm} \iint\limits_{IA,C} d \tilde{p}\hspace{0.05cm} d \tilde{q}\hspace{0.05cm} \tilde{p}\hspace{0.05cm} \tilde{q}^2 (\tilde{p}^2 + \tilde{q}^2) \hspace{0.05cm} e^{-(\tilde{p}^2 + \tilde{q}^2)}\\
&\hspace{-1.86cm} + \iint\limits_{IB,C} d \tilde{p}\hspace{0.05cm} d \tilde{q}\hspace{0.05cm} \tilde{q} \hspace{0.05cm} (\tilde{p}^2 + \tilde{q}^2) \Big(\tilde{\varepsilon}_t - \frac{\tilde{p}^2 + \tilde{q}^2}{2}\Big) \hspace{0.05cm} e^{-(\tilde{p}^2 + \tilde{q}^2)} \Bigg]. \label{cg}
\end{align}

A two-way collision which scatters a particle to the $\vec{n}$-th state increases $E_r$ by $\varepsilon_d - E_n$. Therefore,
\begin{equation}
\bigg(\frac{dE_r}{dt}\bigg)^g_2 = \sum_{\vec{n}} (\varepsilon_d - E_n)  \bigg(\frac{dN_{\vec{n}}}{dt}\bigg)^g_{2}\;.
\label{Erg2app}
\end{equation}

Since the occupation of a dimple state depends only on its energy (Eqs. (\ref{ng1}) and (\ref{ng2})), the dimple population can be described by a continuous distribution function $f(\tilde{E},t) = D(\tilde{E}) N_{\vec{n}} (t) / \mathcal{N}$ with $\tilde{E}_n = \tilde{E}$, where $D(\tilde{E}) = 2 \tilde{l}_d^3 (\tilde{E} / \pi)^{1/2}$ denotes the density of states. Using this definition in Eqs. (\ref{ngapp})$-$(\ref{ng2}) we find the growth rate of $f(\tilde{E},t)$:
\begin{equation}
\bigg(\frac{\partial f(\tilde{E},t)}{\partial t}\bigg)^g = \bigg(\frac{\partial f(\tilde{E},t)}{\partial t}\bigg)^g_{1} + \bigg(\frac{\partial f(\tilde{E},t)}{\partial t}\bigg)^g_{2}\;,
\label{fg}
\end{equation}
where
\begin{align}
\nonumber \bigg(\frac{\partial f(\tilde{E},t)}{\partial t}\bigg)^g_{1} = &\hspace{0.1cm} G \frac{\tilde{z}_r^2}{\tilde{\beta}_r^2} \Big[2\Big(\frac{\tilde{E}}{\pi}\Big)^{\frac{1}{2}} \hspace{-0.05cm}+ \rho_i \Omega f(\tilde{E},t)\Big] \hspace{2cm}\\
&\times \mathscr{G}_1 (\tilde{\beta}_r \tilde{\varepsilon}_t, \tilde{\beta}_r \tilde{\varepsilon}_d, \tilde{\beta}_r \tilde{E}) \;,\label{fg1}\\
\nonumber \bigg(\frac{\partial f(\tilde{E},t)}{\partial t}\bigg)^g_{2} = &\hspace{0.1cm} G \frac{\tilde{z}_r^2}{\tilde{\beta}_r^2} \hspace{0.05cm} \mathscr{G}_2 (\tilde{\beta}_r \tilde{\varepsilon}_t, \tilde{\beta}_r \tilde{\varepsilon}_d, \tilde{\beta}_r \tilde{E}) \\
&\hspace{-2.2cm} \times \Big[2\Big(\frac{\tilde{E}}{\pi}\Big)^{\frac{1}{2}} \hspace{-0.05cm}- \rho_i \Omega f(\tilde{E},t) \Big\{\frac{1}{z_r} e^{-\tilde{\beta}_r (\tilde{\varepsilon}_d - \tilde{E})} - 1 \Big\}\Big],\label{fg2}
\end{align}
with the ``rate constant" $G$ given by
\begin{align}
\nonumber G \equiv \frac{\tilde{G}}{\rho_i \Omega} &= \frac{16\sqrt{2}}{\beta_{r0} \hbar} \Big(\frac{a}{\lambda_{r0}}\Big)^2 \frac{\rho_i}{\Omega} \frac{\pi/4}{(\gamma(3/2,\tilde{\varepsilon}_t))^2}\\
&= \frac{1}{\tau_{\mbox{\scriptsize{coll}}} \Omega} \frac{\sqrt{\pi / 2}}{\gamma(2,\tilde{\varepsilon}_t) \gamma(3/2, \tilde{\varepsilon}_t)}\;.
\label{G}
\end{align}
Here we have substituted from Eqs. (\ref{rhoi}), (\ref{taucoll}), and (\ref{omega}). We note that the characteristic timescale for the growth of the dimple population is $\tau_{\mbox{\scriptsize{coll}}} \Omega$.

Condensation occurs when a macroscopic number of particles reside in the ground state. The condensate fraction is defined as $f_0 (t) \equiv N_{\vec{0}} (t) / \mathcal{N}$. Using Eqs. (\ref{ngapp})$-$(\ref{ng2}) we obtain
\begin{align}
\bigg(\frac{df_{0}}{dt}\bigg)^g_{1} =&\hspace{0.1cm} G \frac{\tilde{z}_r^2}{\tilde{\beta}_r^2} \Big(\frac{1}{\tilde{l}_d^3} + \rho_i \Omega f_{0}\Big) \hspace{0.05cm} \mathscr{G}_1 (\tilde{\beta}_r \tilde{\varepsilon}_t, \tilde{\beta}_r \tilde{\varepsilon}_d, 0) \;,\label{f0g1}\\
\nonumber \bigg(\frac{df_{0}}{dt}\bigg)^g_{2} =&\hspace{0.1cm} G \frac{\tilde{z}_r^2}{\tilde{\beta}_r^2} \bigg[\frac{1}{\tilde{l}_d^3} - \rho_i \Omega f_{0} \Big\{\frac{1}{z_{r}} e^{-\tilde{\beta}_r \tilde{\varepsilon}_d} - 1\Big\}\bigg]\\
& \times \mathscr{G}_2 (\tilde{\beta}_r \tilde{\varepsilon}_t, \tilde{\beta}_r \tilde{\varepsilon}_d, 0) \;.\label{f0g2}
\end{align}

Similarly, Eqs. (\ref{nrgapp}), (\ref{Erg1app2}), and (\ref{Erg2app}) yield the growth rates of the reservoir fraction $f_r \equiv N_r / \mathcal{N}$ and of $\tilde{e}_r \equiv e_r (\gamma(4,\tilde{\varepsilon}_t) / \gamma(3,\tilde{\varepsilon}_t)) = (E_r / \mathcal{E})(\gamma(4,\tilde{\varepsilon}_t) / \gamma(3,\tilde{\varepsilon}_t))$:
\begin{align}
\nonumber \bigg(\frac{df_{r}}{dt}\bigg)^g =&\hspace{0.05cm} -2 \bigg(\frac{df_{0}}{dt}\bigg)^g_{1} - \bigg(\frac{df_{0}}{dt}\bigg)^g_{2} \\
&\hspace{-1cm} - \int_{0}^{\tilde{\varepsilon}_d} d \tilde{E} \bigg[2 \bigg(\frac{\partial f (\tilde{E}, t)}{\partial t}\bigg)^g_{1} + \bigg(\frac{\partial f (\tilde{E}, t)}{\partial t}\bigg)^g_{2}\hspace{0.05cm}\bigg] , \label{frgapp}\\
\nonumber \bigg(\frac{d\tilde{e}_{r}}{dt}\bigg)^g_1 =&\hspace{0.05cm} -G \frac{\tilde{z}_r^2}{\tilde{\beta}_r^3} \bigg[\Big(\frac{1}{\tilde{l}_d^3} + \rho_i \Omega f_{0}\Big) \mathscr{C}_g (\tilde{\beta}_r \tilde{\varepsilon}_t, \tilde{\beta}_r \tilde{\varepsilon}_d, 0) \\
&\hspace{-1.45cm} + \int_{0}^{\tilde{\varepsilon}_d} d \tilde{E} \Big[2\Big(\frac{\tilde{E}}{\pi}\Big)^{\frac{1}{2}} \hspace{-0.05cm}+ \rho_i \Omega f(\tilde{E},t)\Big] \mathscr{C}_g (\tilde{\beta}_r \tilde{\varepsilon}_t, \tilde{\beta}_r \tilde{\varepsilon}_d, \tilde{\beta}_r \tilde{E}) \bigg], \label{erg1}\\
\bigg(\frac{d\tilde{e}_{r}}{dt}\bigg)^g_2 =&\hspace{0.1cm} \tilde{\varepsilon}_d \bigg(\frac{df_{0}}{dt}\bigg)^g_{2} + \int_{0}^{\tilde{\varepsilon}_d} d \tilde{E} \hspace{0.05cm} (\tilde{\varepsilon}_d - \tilde{E})  \bigg(\frac{\partial f(\tilde{E},t)}{\partial t}\bigg)^g_{2} \hspace{0.05cm}.\label{erg2}
\end{align}

\section{Rate equations for the redistribution of the dimple population}\label{redistributionapp}

The rate at which particles are scattered from state $\vec{n}_1$ to state $\vec{n}_2$ of the dimple is given by (Eq. (\ref{n1n2}))
\begin{align}
\nonumber \frac{dN_{\vec{n}_1 \to \vec{n}_2}}{dt} =& \hspace{0.05cm}\frac{2 \pi}{\hbar} U_0^2 \hspace{0.05cm}\frac{z_r}{l_d^3} N_{\vec{n}_1} (1 + N_{\vec{n}_2}) \int^{\prime} \hspace{-0.2cm} \frac{d^3 p}{(2 \pi \hbar)^3} \; e^{-\beta_r \frac{p^2}{2m}}\\
\times \hspace{0.05cm} \delta\bigg(\frac{p^2}{2m} &+ E_{n_1} - \frac{\big(\vec{p} + (2 \pi \hbar / l_d) (\vec{n}_1 - \vec{n}_2)\big)^2}{2m} - E_{n_2}\hspace{-0.05cm}\bigg)\hspace{0.05cm}, \label{n1n2app}
\end{align} 
where the prime restricts the initial energy of the reservoir particle below the trap depth: $p^2 < 2 m \varepsilon_t$. We can write Eq. (\ref{n1n2app}) more simply in terms of $\vec{\tilde{p}} \equiv (\beta_{r}/2m)^{1/2} \vec{p}\hspace{0.05cm}$:
\begin{equation}
\frac{dN_{\vec{n}_1 \to \vec{n}_2}}{dt} = \tilde{R} \frac{\tilde{z}_r}{\tilde{\beta}_r} \frac{N_{\vec{n}_1} (1 + N_{\vec{n}_2}) \int^{\star} \hspace{-0.05cm} d \tilde{p} \hspace{0.08cm} 2 \hspace{0.02cm} \tilde{p} \hspace{0.05cm} e^{-\tilde{p}^2}}{\big(\tilde{E}_{n_1} + \tilde{E}_{n_2} - 2 (\tilde{E}_{n_1} \tilde{E}_{n_2})^{1/2} \hspace{0.05cm}\hat{n}_1.\hat{n}_2 \big)^{1/2}} \hspace{0.05cm},\label{n1n2sub}
\end{equation}
where $\hat{n}_i \equiv \vec{n}_i / n_i$, $\tilde{R} \equiv (4 \pi a^2 z_{r0} / l_d^3) \sqrt{2 / m \beta_{r0}}$, and the asterisk imposes the condition $\beta_r E_{n_2} \cos^2 \theta(\vec{n}_2, \vec{n}_1 - \vec{n}_2) < \tilde{p}^2 < \beta \varepsilon_t$, $\theta(\vec{n}_2, \vec{n}_1 - \vec{n}_2)$ being the angle between $\vec{n}_2$ and $\vec{n}_1 - \vec{n}_2$. The lower limit on $\tilde{p}$ arises from conservation of energy and momentum. When $E_{n_2} < E_{n_1}$, the reservoir particle recoils with a higher energy. If $\tilde{p}^2 > \beta_r (\varepsilon_t - E_{n_1} + E_{n_2})$, its energy exceeds $\varepsilon_t$ and it is lost from the trap. Such collisions have no reverse process. Whereas for $\tilde{p}^2 < \beta_r (\varepsilon_t - E_{n_1} + E_{n_2})$, no particle is lost and collisions happen both ways. When $E_{n_2} > E_{n_1}$, every scattering event which transfers a particle from state $\vec{n}_1$ to state $\vec{n}_2$ can happen backward as well. Thus we can identify the contributions of one-way and two-way collisions in Eq. (\ref{n1n2sub}):
\begin{align}
\nonumber & \bigg(\frac{dN_{\vec{n}_1 \to \vec{n}_2}}{dt} \bigg)_1 = \tilde{R} \frac{\tilde{z}_r}{\tilde{\beta}_r} N_{\vec{n}_1} (1 + N_{\vec{n}_2}) \;\times\\
& \frac{\alpha \big(e^{-\mbox{\scriptsize{max}}\{\tilde{\beta}_r \tilde{E}_{n_2} \cos^2 \hspace{-0.05cm} \theta(\vec{n}_2, \vec{n}_1 - \vec{n}_2), \hspace{0.05cm} \tilde{\beta}_r (\tilde{\varepsilon}_t - \tilde{E}_{n_1} + \tilde{E}_{n_2})\}} - e^{-\tilde{\beta}_r \tilde{\varepsilon}_t}\big)}{\big(\tilde{E}_{n_1} + \tilde{E}_{n_2} - 2 (\tilde{E}_{n_1} \tilde{E}_{n_2})^{1/2} \hspace{0.05cm}\hat{n}_1.\hat{n}_2 \big)^{1/2}} \hspace{0.05cm},\label{n1n21}\\
\nonumber & \bigg(\frac{dN_{\vec{n}_1 \to \vec{n}_2}}{dt} \bigg)_2 = \tilde{R} \frac{\tilde{z}_r}{\tilde{\beta}_r} N_{\vec{n}_1} (1 + N_{\vec{n}_2}) \;\times\\
& \frac{\alpha \big(e^{-\tilde{\beta}_r \tilde{E}_{n_2} \cos^2 \hspace{-0.05cm} \theta(\vec{n}_2, \vec{n}_1 - \vec{n}_2)} - e^{-\mbox{\scriptsize{min}}\{\tilde{\beta}_r \tilde{\varepsilon}_t, \hspace{0.05cm} \tilde{\beta}_r (\tilde{\varepsilon}_t - \tilde{E}_{n_1} + \tilde{E}_{n_2})\}}\big)}{\big(\tilde{E}_{n_1} + \tilde{E}_{n_2} - 2 (\tilde{E}_{n_1} \tilde{E}_{n_2})^{1/2} \hspace{0.05cm}\hat{n}_1.\hat{n}_2 \big)^{1/2}} \hspace{0.05cm},\label{n1n22}
\end{align}
where $\alpha$ denotes the ramp function: $\alpha(x) = x$ for $x>0$, and $\alpha(x) = 0$ for $x<0$. The overall rate of change of $N_{\vec{n}}$ due to particle transfer from other states can then be written as
\begin{equation}
\bigg(\frac{dN_{\vec{n}}}{dt}\bigg)^r = \sum_{\vec{n}^{\prime} \neq \vec{n}} R^{(1)}_{\vec{n},\vec{n}^{\prime}} + R^{(2)}_{\vec{n},\vec{n}^{\prime}}\;, \label{nnr12app}
\end{equation}
where
\begin{equation}
R^{(i)}_{\vec{n},\vec{n}^{\prime}} \equiv \bigg(\frac{dN_{\vec{n}^{\prime} \to \vec{n}}}{dt} \bigg)_i - \bigg(\frac{dN_{\vec{n} \to \vec{n}^{\prime}}}{dt} \bigg)_i\hspace{0.05cm}, \quad i = 1,2.
\label{Rinnprime}
\end{equation}

Two-way collisions do not alter the number of particles in the reservoir ($N_r$), whereas each one-way collision removes one particle from the reservoir. Therefore,
\begin{equation}
\bigg(\frac{dN_r}{dt}\bigg)^r = -\hspace{-0.1cm}\sum_{\substack{\vec{n}^{\prime}, \vec{n} \\ E_{n^{\prime}} > E_n}} \hspace{-0.1cm} R^{(1)}_{\vec{n},\vec{n}^{\prime}}\;. \label{nrrapp}
\end{equation}

In a one-way collision, the reservoir particle is lost from the trap. This reduces the energy in the reservoir ($E_r$) by $p^2 / 2m$. The expression for the net rate of decrease of $E_r$ looks similar to Eq. (\ref{n1n2app}) and is given in Eq. (\ref{Errn1n2}). To simplify we perform the same substitutions as in Eq. (\ref{n1n2app}), thus obtaining
\begin{align}
\nonumber & \bigg(\frac{dE_r}{dt} \bigg)_1^r = -\tilde{R} \frac{\tilde{z}_r}{\beta_r \tilde{\beta}_r} \sum_{\substack{\vec{n}^{\prime}, \vec{n} \\ E_{n^{\prime}} > E_n}} N_{\vec{n}^{\prime}} (1 + N_{\vec{n}}) \;\times\\
& \frac{\xi \big(\mbox{max}\{\tilde{\beta}_r \tilde{E}_{n} \cos^2 \hspace{-0.05cm} \theta(\vec{n}, \vec{n}^{\prime} - \vec{n}), \tilde{\beta}_r (\tilde{\varepsilon}_t - \tilde{E}_{n^{\prime}} + \tilde{E}_{n})\}, \tilde{\beta}_r \tilde{\varepsilon}_t\big)}{\big(\tilde{E}_{n} + \tilde{E}_{n^{\prime}} - 2 (\tilde{E}_{n} \tilde{E}_{n^{\prime}})^{1/2} \hspace{0.05cm}\hat{n}.\hat{n}^{\prime} \big)^{1/2}} \hspace{0.03cm},\label{Err1simplified}
\end{align}
where $\xi (a,b) \equiv \big((1+a) \hspace{0.05cm} e^{-a} - (1+b) \hspace{0.05cm} e^{-b}\big) \hspace{0.05cm} \Theta(b-a)$, $\Theta$ being the Heaviside step function: $\theta(x) = 1$ for $x>0$, and $\theta(x) = 0$ for $x<0$.

A two-way collision which transfers a particle from state $\vec{n}^{\prime}$ to state $\vec{n}$ increases $E_r$ by $E_{n^{\prime}} - E_n$. Hence,
\begin{equation}
\bigg(\frac{dE_r}{dt}\bigg)^r_2 = \sum_{\substack{\vec{n}^{\prime}, \vec{n} \\ E_{n^{\prime}} > E_n}} \hspace{-0.1cm} (E_{n^{\prime}} - E_n) \hspace{0.05cm} R^{(2)}_{\vec{n},\vec{n}^{\prime}}\hspace{0.05cm}.\label{Err2app}
\end{equation}

Due to symmetry, $N_{\vec{n}}$ depends only on $E_n$ (Eqs. (\ref{n1n21})$-$(\ref{Rinnprime})). Thus we describe the discrete states in the dimple by a continuous density of states $D(\tilde{E}) = 2 \tilde{l}_d^3 (\tilde{E} / \pi)^{1/2}$ and their occupations by a distribution function $f(\tilde{E},t) \equiv D(\tilde{E}) N_{\vec{n}}(t) / \mathcal{N}$ where $\tilde{E}_n = \tilde{E}$. Then Eqs. (\ref{n1n21})$-$(\ref{Rinnprime}) give
\begin{equation}
\hspace{-0.18cm}\bigg(\frac{\partial f(\tilde{E},t)}{\partial t}\bigg)^r = \int_{0}^{\tilde{\varepsilon}_d} d \tilde{E^{\prime}} \big(\mathcal{R}_1 (\tilde{E}, \tilde{E}^{\prime}) + \mathcal{R}_2 (\tilde{E}, \tilde{E}^{\prime})\big) \hspace{0.05cm},\label{frapp}
\end{equation}
where $\mathcal{R}_i (\tilde{E}^{\prime}, \tilde{E}) = - \mathcal{R}_i (\tilde{E}, \tilde{E}^{\prime})$ and for $\tilde{E}^{\prime} > \tilde{E}$,
\begin{align}
\nonumber \mathcal{R}_1 (\tilde{E}, \tilde{E}^{\prime}) \equiv &\hspace{0.1cm} R \frac{\tilde{z}_r}{\tilde{\beta}_r^{1/2}} f(\tilde{E}^{\prime},t) \Big[2\Big(\frac{\tilde{E}}{\pi}\Big)^{\frac{1}{2}} \hspace{-0.05cm}+ \rho_i \Omega f(\tilde{E},t)\Big] \\
&\hspace{0.1cm} \times \mathcal{I}_{A}^{(1)} (\tilde{\beta}_r \tilde{\varepsilon}_t, \tilde{\beta}_r \tilde{E}, \tilde{\beta}_r \tilde{E}^{\prime})\hspace{0.05cm},\label{R1}\\
\nonumber \mathcal{R}_2 (\tilde{E}, \tilde{E}^{\prime}) \equiv &\hspace{0.1cm} R \frac{\tilde{z}_r}{\tilde{\beta}_r^{1/2}} \bigg[f(\tilde{E}^{\prime},t) \Big\{2\Big(\frac{\tilde{E}}{\pi}\Big)^{\frac{1}{2}} \hspace{-0.05cm}+ \rho_i \Omega f(\tilde{E},t)\Big\}\\
\nonumber &\hspace{-1cm} -f(\tilde{E},t) \Big\{2\Big(\frac{\tilde{E}^{\prime}}{\pi}\Big)^{\frac{1}{2}} \hspace{-0.05cm}+ \rho_i \Omega f(\tilde{E}^{\prime},t)\Big\} \hspace{0.05cm} e^{-\tilde{\beta}_r (\tilde{E}^{\prime} - \tilde{E})}\bigg] \\
& \hspace{0.1cm} \times \mathcal{I}_{A}^{(2)} (\tilde{\beta}_r \tilde{\varepsilon}_t, \tilde{\beta}_r \tilde{E}, \tilde{\beta}_r \tilde{E}^{\prime})\hspace{0.05cm},\label{R2}
\end{align}
where the ``rate constant" $R$ has the expression
\begin{equation}
R = \frac{2 \sqrt{\pi}}{\beta_{r0} \hbar} \Big(\frac{a}{\lambda_{r0}}\Big)^2 \rho_i \frac{\sqrt{\pi}/2}{\gamma(3/2,\tilde{\varepsilon}_t)} = \frac{\sqrt{\pi} \hspace{0.05cm} \gamma(2,\tilde{\varepsilon}_t)}{8 \hspace{0.05cm} \tau_{\mbox{\scriptsize{coll}}}} \hspace{0.05cm},\label{R}
\end{equation}
and the functions $\mathcal{I}_{A}^{(1)}$ and $\mathcal{I}_{A}^{(2)}$ are given by
\begin{align}
\nonumber &\mathcal{I}_{A}^{(1)} (\tilde{\varepsilon}_t, \tilde{E}, \tilde{E}^{\prime}) \equiv \\
&\hspace{0.5cm}\int_{-1}^{1} du \hspace{0.05cm} \frac{\alpha \Big(e^{-\mbox{\scriptsize{max}} \big\{ \tilde{E} \frac{(\sqrt{\tilde{E}^{\prime}}u - \sqrt{\tilde{E}})^2}{\tilde{E} + \tilde{E}^{\prime} - 2 \sqrt{\tilde{E} \tilde{E}^{\prime}} u}, \hspace{0.05cm} \tilde{\varepsilon}_t - \tilde{E}^{\prime} + \tilde{E}\big\}} - e^{-\tilde{\varepsilon}_t}\Big)}{\big(\tilde{E} + \tilde{E}^{\prime} - 2 \sqrt{\tilde{E} \tilde{E}^{\prime}} \hspace{0.03cm} u\big)^{1/2}} \hspace{0.05cm},\label{ia1}\\
\nonumber &\mathcal{I}_{A}^{(2)} (\tilde{\varepsilon}_t, \tilde{E}, \tilde{E}^{\prime}) \equiv\\
& \hspace{0.5cm} \int_{-1}^{1} du \hspace{0.05cm} \frac{\alpha \Big(e^{-\tilde{E} \frac{(\sqrt{\tilde{E}^{\prime}}u - \sqrt{\tilde{E}})^2}{\tilde{E} + \tilde{E}^{\prime} - 2 \sqrt{\tilde{E} \tilde{E}^{\prime}} u}} - e^{-(\tilde{\varepsilon}_t - \tilde{E}^{\prime} + \tilde{E})}\Big)}{\big(\tilde{E} + \tilde{E}^{\prime} - 2 \sqrt{\tilde{E} \tilde{E}^{\prime}} \hspace{0.03cm} u\big)^{1/2}} \hspace{0.05cm}.\label{ia2}
\end{align}
We note that the characteristic timescale in Eqs. (\ref{R1}) and (\ref{R2}) is $\tau_{\mbox{\scriptsize{coll}}}$. Thus we expect $\tau_{\mbox{\scriptsize{coll}}}$ to set the thermalization time scale.

The equation of motion for the condensate fraction $f_0 (t) \equiv N_{\vec{0}}(t) / \mathcal{N}$ can be obtained likewise:
\begin{align}
\nonumber \bigg(\frac{df_{0}}{dt}\bigg)^r_1 =&\hspace{0.1cm} R \frac{\tilde{z}_r}{\tilde{\beta}_r^{1/2}} \Big(\frac{1}{\tilde{l}_d^3} + \rho_i \Omega f_{0}\Big)\\ 
& \hspace{0.05cm} \times \int_0^{\tilde{\varepsilon}_d} d \tilde{E} f (\tilde{E},t) \hspace{0.05cm} \mathcal{I}_{A}^{(1)} (\tilde{\beta}_r \tilde{\varepsilon}_t, 0, \tilde{\beta}_r \tilde{E})\;,\label{f0r1}\\
\bigg(\frac{df_{0}}{dt}\bigg)^r_2 =& \int_0^{\tilde{\varepsilon}_d} d \tilde{E} \hspace{0.05cm} \mathcal{R}_0 (\tilde{E}) \;,\label{f0r2}
\end{align}
where
\begin{align}
\nonumber \mathcal{R}_0 (\tilde{E}) \equiv &\hspace{0.1cm} R \frac{\tilde{z}_r}{\tilde{\beta}_r^{1/2}} \bigg[f (\tilde{E},t) \Big(\frac{1}{\tilde{l}_d^3} + \rho_i \Omega f_{0}\Big) \\
& \hspace{-1.2cm} -f_0 \Big\{2\Big(\frac{\tilde{E}}{\pi}\Big)^{\frac{1}{2}} \hspace{-0.05cm}+ \rho_i \Omega f(\tilde{E},t)\Big\} \hspace{0.05cm} e^{-\tilde{\beta}_r \tilde{E}} \bigg] \mathcal{I}_{A}^{(2)} (\tilde{\beta}_r \tilde{\varepsilon}_t, 0, \tilde{\beta}_r \tilde{E})\hspace{0.05cm}.\label{R0}
\end{align}
Similarly, the continuum limit of Eqs. (\ref{nrrapp})$-$(\ref{Err2app}) yields
\begin{align}
&\bigg(\frac{df_{r}}{dt}\bigg)^r = -\bigg(\frac{df_{0}}{dt}\bigg)^r_1 - \int_0^{\tilde{\varepsilon}_d} d \tilde{E} \int_{\tilde{E}}^{\tilde{\varepsilon}_d} d \tilde{E}^{\prime} \hspace{0.05cm} \mathcal{R}_1 (\tilde{E}, \tilde{E}^{\prime})\hspace{0.05cm},\label{frr}\\
&\nonumber \bigg(\frac{d\tilde{e}_{r}}{dt}\bigg)^r_1 = -R \frac{\tilde{z}_r}{\tilde{\beta}_r^{3/2}} \bigg[\Big(\frac{1}{\tilde{l}_d^3} + \rho_i \Omega f_{0}\Big) \mathscr{C}_r (\tilde{\beta}_r, \tilde{\varepsilon}_t, 0, t)\\
&+ \int_0^{\tilde{\varepsilon}_d} d \tilde{E} \Big\{2\Big(\frac{\tilde{E}}{\pi}\Big)^{\frac{1}{2}} \hspace{-0.05cm}+ \rho_i \Omega f(\tilde{E})\Big\} \mathscr{C}_r (\tilde{\beta}_r, \tilde{\varepsilon}_t, \tilde{E}, t) \bigg],\label{err1}\\
&\bigg(\frac{d\tilde{e}_{r}}{dt}\bigg)^r_2 = \hspace{-0.05cm}\int_0^{\tilde{\varepsilon}_d} \hspace{-0.1cm} d \tilde{E}^{\prime} \Big[\tilde{E}^{\prime} \mathcal{R}_0 (\tilde{E}^{\prime}) + \hspace{-0.05cm} \int_0^{\tilde{E}^{\prime}} \hspace{-0.1cm}(\tilde{E}^{\prime} - \tilde{E}) \mathcal{R}_2 (\tilde{E}, \tilde{E}^{\prime})\Big],\label{err2}
\end{align}
where $\tilde{e}_{r} \equiv e_r \big(\gamma(4, \tilde{\varepsilon}_t) / \gamma(3, \tilde{\varepsilon}_t) \big)$ (Eq. (\ref{er})) and
\begin{align}
\nonumber &\mathscr{C}_r (\tilde{\beta}_r, \tilde{\varepsilon}_t, \tilde{E}, t) \equiv \int_{\tilde{E}}^{\tilde{\varepsilon}_d} d \tilde{E}^{\prime} f(\tilde{E}^{\prime}, t) \;\times \\
&\int_{-1}^{1} du \hspace{0.02cm}\frac{\xi \Big(\hspace{-0.02cm} \tilde{\beta}_r \mbox{max} \Big\{ \tilde{E} \frac{(\sqrt{\tilde{E}^{\prime}}u - \sqrt{\tilde{E}})^2}{\tilde{E} + \tilde{E}^{\prime} - 2 \sqrt{\tilde{E} \tilde{E}^{\prime}} u}, \hspace{0.05cm} \tilde{\varepsilon}_t - \tilde{E}^{\prime} + \tilde{E}\Big\}, \tilde{\beta}_r \tilde{\varepsilon}_t \hspace{-0.02cm} \Big)}{\tilde{\beta}_r^{1/2} \big(\tilde{E} + \tilde{E}^{\prime} - 2 \sqrt{\tilde{E} \tilde{E}^{\prime}} \hspace{0.03cm} u\big)^{1/2}} \hspace{0.03cm}.\label{cr}
\end{align}

\section{Rate equations for three-body loss}\label{threebodyapp}

Because of three-body loss the overall particle density decays as (Eq. (\ref{densitylossrate2})) \cite{davis11, pethick, kagan85}
\begin{align}
\nonumber \bigg(\frac{dn(\vec{r})}{dt}\bigg)^l = -L \Big[& n_0^3 (\vec{r}) + 9 \hspace{0.05cm} n_0^2 (\vec{r}) \hspace{0.05cm} n_{ex} (\vec{r}) \\
&+ 18 \hspace{0.05cm} n_0 (\vec{r}) \hspace{0.05cm} n_{ex}^2 (\vec{r}) + 6 n_{ex}^3 (\vec{r}) \Big] \hspace{0.05cm},
\label{densitylossrate2app}
\end{align}
where $n_0 (\vec{r})$ and $n_{ex} (\vec{r})$ are the densities of the condensate and the excited-state atoms respectively, and $L$ denotes the loss coefficient which for ${}^{87}$Rb equals $L = 1.8 \times 10^{-29}$ $\mbox{cm}^6 \hspace{0.05cm} \mbox{s}^{-1}$ \cite{dalibard99}. The exponents in Eq. (\ref{densitylossrate2app}) arise from the number of atoms participating in the recombination process: the first term results from recombination of three condensate atoms, whereas the second term originates from events where two condensate atoms recombine with a higher-energy atom etc. Further, an excited-state atom can either be a member of the non-condensate population in the dimple or reside in the reservoir. Thus $n_{ex} (\vec{r}) = n_{nc} (\vec{r}) + n_{r} (\vec{r})$ with $n_{r} (\vec{r})$ given in Eq. (\ref{nr}). Keeping these in mind we write down the decay rates of the individual densities
\begin{eqnarray}
\bigg(\frac{dn_0}{dt}\bigg)^l &=& -L \hspace{0.05cm} n_0 \big(n_0^2 + 6 n_0 n_{ex} + 6 n_{ex}^2 \big)\hspace{0.05cm}, \label{n0lossrate}\\
\bigg(\frac{dn_{nc}}{dt}\bigg)^l &=& -3 \hspace{0.05cm} L \hspace{0.05cm} n_{nc} \big( n_0^2 + 4 n_0 n_{ex} + 2 n_{ex}^2 \big)\hspace{0.05cm}, \label{nnclossrate}\\
\bigg(\frac{dn_{r}}{dt}\bigg)^l &=& -3 \hspace{0.05cm} L \hspace{0.05cm} n_{r} \big( n_0^2 + 4 n_0 n_{ex} + 2 n_{ex}^2 \big)\hspace{0.05cm}. \label{nrlossrate}
\end{eqnarray}
We model the densities in the dimple as $n_0 (\vec{r}) = \mathcal{N} f_0 / l_d^3 = (\rho_i \Omega / \lambda_{r0}^3) f_0$ and $n_{nc} (\vec{r}) = \mathcal{N} f_{nc} / l_d^3 = (\rho_i \Omega / \lambda_{r0}^3) f_{nc}$. Outside the dimple we take $n_0 (\vec{r}) = n_{nc} (\vec{r}) = 0$. In addition, since the dimple is many times smaller than the reservoir, the reservoir density can be taken as uniform inside the dimple: $n_r (\vec{r}) \approx n_r (\vec{0}) = (\rho_i \tilde{z}_r / \lambda_{r0}^3 \tilde{\beta}_r^{3/2}) (\gamma(3/2, \tilde{\beta}_r \tilde{\varepsilon}_t) / \gamma(3/2, \tilde{\varepsilon}_t))$ for points inside the dimple (Eqs. (\ref{nr}) and (\ref{rhoi})). Using these expressions in Eqs. (\ref{n0lossrate}) and (\ref{nnclossrate}) we find
\begin{eqnarray}
\bigg(\frac{df_0}{dt}\bigg)^l &=& -L \frac{\rho_i^2 \Omega^2}{\lambda_{r0}^6} f_0 (f_0^2 + 6 f_0 f^{\prime} + 6 f^{\prime 2})\hspace{0.05cm}, \label{f0lossrate}\\
\bigg(\frac{df_{nc}}{dt}\bigg)^l &=& -3 L \frac{\rho_i^2 \Omega^2}{\lambda_{r0}^6} f_{nc} (f_0^2 + 4 f_0 f^{\prime} + 2 f^{\prime 2})\hspace{0.05cm}, \label{fnclossrate}
\end{eqnarray}
where $f^{\prime} \equiv f_{nc} + (\tilde{z}_r / \Omega \tilde{\beta}_r^{3/2}) (\gamma(3/2, \tilde{\beta}_r \tilde{\varepsilon}_t) / \gamma(3/2, \tilde{\varepsilon}_t))$. In terms of the particle distribution $f(\tilde{E}, t)$ in the excited states of the dimple, $f_{nc} = \int_0^{\tilde{\varepsilon}_d} d \tilde{E} f(\tilde{E},t)$. Under the assumption that double or higher occupancy of the excited states in negligible, it follows from Eq. (\ref{fnclossrate}) that
\begin{equation}
\hspace{-0.2cm}\bigg(\frac{\partial f(\tilde{E},t)}{\partial t}\bigg)^l = -3 L \frac{\rho_i^2 \Omega^2}{\lambda_{r0}^6} f(\tilde{E},t) (f_0^2 + 4 f_0 f^{\prime} + 2 f^{\prime 2})\hspace{0.05cm}. \label{flossrate}
\end{equation}

The three-body decay rate of the reservoir fraction $f_r = (1/\mathcal{N}) \int d^3 r \hspace{0.05cm} n_r (\vec{r})$ can be obtained by substituting $n_r (\vec{r})$ from Eq. (\ref{nr}) and the above expressions for $n_0$ and $n_{nc}$ into Eq. (\ref{nrlossrate}). This yields
\begin{align}
\nonumber &\bigg(\frac{df_{r}}{dt}\bigg)^l = -3 L \frac{\rho_i^2 \tilde{z}_r \Omega}{\lambda_{r0}^6 \tilde{\beta}_r^{3/2}} \frac{\gamma(3/2, \tilde{\beta}_r \tilde{\varepsilon}_t)}{\gamma(3/2, \tilde{\varepsilon}_t)} \bigg(\hspace{-0.05cm} f_0^2 + 4 f_0 f^{\prime} + 2 f_{nc}^{2} \\
\nonumber &+ \hspace{0.05cm} 4 \frac{f_{nc} \tilde{z}_r}{\Omega \tilde{\beta}_r^{3/2}} \frac{\gamma(3/2, \tilde{\beta}_r \tilde{\varepsilon}_t)}{\gamma(3/2, \tilde{\varepsilon}_t)}\bigg) - \frac{48 L}{\pi \gamma(3,\tilde{\varepsilon}_t) (\gamma(3/2,\tilde{\varepsilon}_t))^2} \frac{\rho_i^2 \tilde{z}_r^3}{\lambda_{r0}^6 \tilde{\beta}_r^6} \\
& \hspace{1.3cm} \times \int_0^{\tilde{\beta}_r \tilde{\varepsilon}_t} \hspace{-0.05cm} dx \hspace{0.05cm} \sqrt{x} \hspace{0.05cm} e^{-3 x} (\gamma(3/2, \tilde{\beta}_r \tilde{\varepsilon}_t - x))^3 \hspace{0.05cm}.\label{frlossrate}
\end{align}

The total energy in the reservoir ($E_r$) also decays because of three-body loss. We can find the decay rate as $(dE_r/dt)^l = - \int \hspace{-0.05cm} d^3 r \hspace{0.03cm} u_r (\vec{r}) (d n_r / dt)^l$, where $u_r (\vec{r})$ denotes the average energy of a reservoir particle at position $\vec{r}$. To calculate $u_r (\vec{r})$ we integrate over the phase space, finding
\begin{equation}
u_r (\vec{r}) = \frac{1}{\beta_r} \frac{\gamma(5/2, \beta_r \varepsilon_t - \beta_r m \omega^2 r^2 / 2)}{\gamma(3/2, \beta_r \varepsilon_t - \beta_r m \omega^2 r^2 / 2)} + \frac{1}{2} m \omega^2 r^2 \hspace{0.05cm}.\label{ur}
\end{equation}
When $\varepsilon_t \to \infty$, this reduces to the familiar expression $(3/2)k_B T_r + (1/2) m \omega^2 r^2$. Using Eqs. (\ref{nrlossrate}) and (\ref{ur}) we obtain the decay rate of $\tilde{e}_r \equiv (E_r/\mathcal{E}) (\gamma(4,\tilde{\varepsilon}_t) / \gamma(3,\tilde{\varepsilon}_t))$ as
\begin{align}
\nonumber &\bigg(\frac{d\tilde{e}_{r}}{dt}\bigg)^l = -3 L \frac{\rho_i^2 \tilde{z}_r \Omega}{\lambda_{r0}^6 \tilde{\beta}_r^{5/2}} \frac{\gamma(5/2, \tilde{\beta}_r \tilde{\varepsilon}_t)}{\gamma(3/2, \tilde{\varepsilon}_t)} \bigg(\hspace{-0.05cm} f_0^2 + 4 f_0 f^{\prime} + 2 f_{nc}^{2} \\
\nonumber &+ \hspace{0.05cm} 4 \frac{f_{nc} \tilde{z}_r}{\Omega \tilde{\beta}_r^{3/2}} \frac{\gamma(3/2, \tilde{\beta}_r \tilde{\varepsilon}_t)}{\gamma(3/2, \tilde{\varepsilon}_t)}\bigg) - \frac{48 L}{\pi \gamma(3,\tilde{\varepsilon}_t) (\gamma(3/2,\tilde{\varepsilon}_t))^2} \frac{\rho_i^2 \tilde{z}_r^3}{\lambda_{r0}^6 \tilde{\beta}_r^7} \\
& \int_0^{\tilde{\beta}_r \tilde{\varepsilon}_t} \hspace{-0.25cm} dx \sqrt{x} e^{-3 x} (\gamma(3/2, \tilde{\beta}_r \tilde{\varepsilon}_t - x))^3 \hspace{-0.05cm}\bigg[\hspace{-0.05cm}\frac{\gamma(5/2, \tilde{\beta}_r \tilde{\varepsilon}_t - x)}{\gamma(3/2, \tilde{\beta}_r \tilde{\varepsilon}_t - x)} + x \hspace{-0.02cm}\bigg]\hspace{-0.05cm}.\label{erlossrate}
\end{align}

\end{document}